\documentclass[12pt,authoryear]{elsarticle}
\usepackage[bottom = 30mm, right = 30mm, top = 30mm, left=30mm]{geometry}
\usepackage{setspace} 
\usepackage[english]{babel}
\usepackage[utf8]{inputenc}
\usepackage[T1]{fontenc}
\usepackage[colorlinks]{hyperref}
\hypersetup{colorlinks,
	linkcolor={black},
	citecolor={black},
	urlcolor={black}}
\usepackage{indentfirst}
\usepackage{comment}
\usepackage{natbib}
\usepackage{lscape}
\usepackage{amsmath}
\usepackage{booktabs}
\usepackage{threeparttable}
\usepackage{changepage}
\usepackage{amssymb}
\usepackage{array}
\usepackage{wrapfig}
\usepackage{subcaption}
\usepackage{enotez}
\usepackage{changepage}
\usepackage{pgfplots}
\usepackage{floatrow}
\newfloatcommand{capbtabbox}{table}[][\FBwidth]
\usepackage{longtable}
\usepackage{lscape}
\newtheorem{definition}{Definition}
\usepackage{tikz}
\usepackage{pgflibrarysnakes}
\usetikzlibrary{positioning,shapes,backgrounds, decorations.pathmorphing}
\usepackage{float}
\usepackage[singlelinecheck=false]{caption}
\usepackage[edges]{forest}

\begin{document}
\begin{frontmatter}

\title{The geometry of conflict : 3D Spatio-temporal patterns in fatalities prediction}
\author{Thomas Schincariol} 
\ead{schincat@tcd.ie}
\affiliation{organization={Trinity College Dublin},
            addressline={College Green, Dublin 2}, 
            city={Dublin},
            postcode={D02PN40},
            country={Ireland}}

\begin{abstract}
Understanding how conflict events spread over time and space is crucial for predicting and mitigating future violence. However, progress in this area has been limited by the lack of methods capable of capturing the intricate, dynamic patterns of conflict diffusion. The complex nature of those trends needs flexibility in the models to untangle them. This study addresses this gap by analyzing spatio-temporal conflict fatality data using an innovative approach that transforms the data into three-dimensional patterns at the Prio-Grid level. In this paper, a shape-based model called ShapeFinder is adapted. By applying the Earth Mover’s Distance (EMD) algorithm, we detect and classify these patterns, allowing us to compare and match patterns with high adaptive capacity in all dimensions. Using historical similar patterns, we generate predictions of conflict fatalities and compare these with forecasts from the Views ensemble model, a leading benchmark. Our findings demonstrate that recognizing and analyzing conflict diffusion patterns significantly improves predictive accuracy, outperforming the benchmark model. This research contributes to the study of conflict dynamics by introducing a novel pattern recognition framework that enhances the analysis of spatio-temporal data and offers practical applications for early warning systems.
\end{abstract}


\end{frontmatter}

\doublespacing
\clearpage
\section{Introduction}

The spatio-temporal patterns of conflict have been a central focus for scholars over the past few decades. Understanding the conflict dynamics mechanisms could enhance the performance of fatalities forecasting to improve early warning systems further. Research has shown that one of the most reliable predictors of future fatalities is the lagged values of fatalities in the unit of analysis and its neighboring areas. However, clearly identifying the conflict underlying structures remains a challenge. Furthermore, most studies have primarily focused on civil conflict \citep{forsberg2014diffusion} or specific cases \citep{saideman2012conflict, kibris2021geo}. 

Conflict mechanisms are often studied through the lens of diffusion, where fatalities are seen to spread across regions, like an epidemic. Diffusion is mostly studied as a general phenomenon occurring similarly across space and time. However, studying the specific patterns observed in conflict data might bring different information that expands our understanding of conflict dynamics and consequently improves the conflict forecast accuracy. Forecasting models rely on past data to predict future events. Focusing only on historically similar cases can help distinguish valuable information that strengthens forecasting accuracy from unrelated dynamics that introduce noise.  For example, \cite{schutte2011diffusion} identifies distinct patterns in the spread of conflict across four civil war cases. Similarly, \cite{kelling2020analysis}, using the case of South Sudan, shows how analyzing internal diffusion patterns can deepen our understanding of war dynamics.

The generalizability of this approach is the next potential goal of researchers. Most traditional approaches rely on regression models \citep{racek2025capturing}, often related to the Hawkes process \citep{hawkes1971spectra}, where fatalities in a given location increase the likelihood of subsequent conflict events in the neighbouring area. These models produce valuable insights but lack the flexibility needed to disentangle and compare the spatio-temporal patterns of conflicts. The mechanisms underlying conflict are intrinsically complex, requiring models capable of capturing events that, depending on their environment, unfold at varying speeds or scales. Current methods often perform a sophisticated averaging of neighbors, failing to identify specific spatio-temporal patterns. While some machine learning methods offer the potential for the flexibility needed to capture these complex dynamics, their application is restricted by significant drawbacks. These include their dependency on vast amounts of data and limited transparency, which restricts their capacity to generate clear explanations of findings, and often tend to underperform on tabular data \citep{shwartz2022tabular,fayaz2022deep}.

This paper introduces a novel methodology designed to capture complex spatio-temporal conflict patterns, called `ShapeFinder'. Originally developed for country-level conflict forecasting \citep{schincariol2025accounting}, the framework is adapted here to accommodate three-dimensional spatio-temporal patterns. The use of the distance method called Earth Mover's Distance (EMD) gives the model a certain flexibility, enabling the comparison and matching of patterns across different geographic scales and intensities. This approach advances our understanding of conflict mechanisms by identifying significant patterns and contributing to making prediction models more accurate. A notable benefit of this autoregressive model is its simplicity and speed. It does not require any additional variables, which makes it easier to set up and use. The primary strength of the model is its transparency. By leveraging similar patterns from historical data to construct potential scenarios, it becomes possible to trace the cases driving predictions. The outcome information is twofold. First, the potential future scenarios with their occurrence in the historical data provide information about what happened in history when this pattern occurred. Second, these historical matches allow a deeper analysis of the context in which this pattern emerges. This approach enables the association of specific dynamics with particular regions or time periods. This model contributes by offering an accessible spatio-temporal model that captures recurring patterns and uses them to forecast. This model holds a significant interest and has a distinct place in the forecasting literature, particularly when transparency is needed and when the data is highly zero-inflated and skewed, like the conflict fatalities data. 

Transparency in forecasts is essential for convincing non-expert audiences of a model’s credibility and for encouraging decision-makers and NGOs to adopt it as an early warning tool. ACLED’s Conflict Alert System (CAST) \citep{acled_cast_2025}, for example, includes a “What’s driving the forecasts?” feature that identifies the key variables influencing each prediction. Similarly, the ConflictForecast team \citep{mueller2024introducing} presents the topics that increase or decrease predicted conflict risk of their text-based conflict forecasting model. In this context, the simplicity and transparency of the ShapeFinder are important advantages. Additionally, the model’s forecasting performance is evaluated against the Views ensemble model \citep{hegre_forecasting_2022}, one of the leading benchmarks, using both fine-grained accuracy at the Prio-Grid level and broader assessments in larger zones of aggregated cells with both overall accuracy and dynamic fidelity. The ShapeFinder delivers convincing results on both the local and aggregate levels while maintaining transparency, making it a credible forecasting tool.

The evaluation of conflict models at fine-grained geographic scales is complex and remains a subject of debate within the conflict forecasting community. Prio-Grid is widely used below the country level but introduces several challenges, mostly linked with the uncertainty in reported conflict events \citep{weidmann2015accuracy}. At the same time, the prediction in point estimate is increasingly being replaced by probability distributions, which provide users with essential information about forecast uncertainty, essential for a better understanding of the risk associated. This shift introduces additional evaluation metrics, such as the Continuous Ranked Probability Score (CRPS), the probabilistic analog of mean absolute error, and uncertainty measures like the Mean Interval Score (MIS), which rewards narrower prediction intervals. More broadly, questions regarding the appropriate unit of analysis and the most suitable evaluation metrics would require a dedicated study. Here, the scope of the paper is intentionally centered around the model presentation and evaluation with classical metrics. However, flexibility remains regarding both the unit of analysis and the prediction format, and potential adaptations are discussed in the Discussion section. 

Such a model could be valuable for social science research, and particularly for conflict studies, by enabling deeper analysis of conflict patterns. Researchers could investigate the recurrence of specific patterns and their subsequent trajectories to determine whether certain configurations are associated with escalating intensity or the spatial spread of instability. The model could also be used to assess the impact of a given treatment by comparing outcomes across similar patterns that either received the treatment or did not, allowing researchers to study effects both over time and across space.

The remainder of the paper is organized as follows. The next section reviews the relevant literature, followed by a description of the data selection and pre-processing steps. The model methodology is then detailed, together with the benchmark evaluation setup. The results are presented at both the Prio-Grid level and the active-zone level. The paper concludes with a discussion of the model’s limitations and potential future adaptations.

\section{Conflict Forecasting at the subnational level}

Classical spatio-temporal models were originally developed in fields such as epidemiology, ecology, and traffic or urban planning. Deep learning approaches, including Graph Neural Networks \citep{corso2024graph} and Convolutional Networks \citep{yu2017spatio}, have gained popularity in recent years for their strong predictive performance, but they still lack interpretability and simplicity, combined with high training data needs. Models such as self-exciting point processes \citep{reinhart2018review}, derived from Hawkes processes, offer better interpretability and have been applied to rare-event domains like crime risk \citep{reinhart2018self} and seismology \citep{davis2024fractional}. Similarly, the Spatiotemporal Autoregressive Distributed Lag (STADL) model \citep{cook2023stadl} provides clear and interpretable insights on spatiotemporal dependencies. However, the nature of conflict data, characterized by context-driven, time-varying dynamics and abrupt shifts, limits the effectiveness of these approaches. 

Yet, research over the past decades suggests that conflict often spreads across space and time \citep{black2013have,kathman2010civil}, especially at the subnational level. Similar to how diseases spread, conflicts can be modeled using diffusion or network models \citep{dorussen2016networked}. Studies show that if neighboring regions experience conflict, nearby areas are also at risk in the future \citep{buhaug2008contagion}. As a result, using lagged time and spatial data has become essential for forecasting models \citep{hegre2019views}. Besides autoregressive factors, variables such as GDP, population, ethnic exclusion \citep{metternich2017firewall}, political instability \citep{iqbal2008bad}, and variables extracted from new sources like remote sensors \citep{racek2024conflict} also contribute to explaining conflict.

To forecast conflict at a regional level, researchers have used advanced statistical techniques \citep{fritz2022role}, but also machine learning approaches \citep{bazzi2022promise} are increasingly common. Many of these models include spatial and temporal data by incorporating lagged values from neighboring areas using space and time-lagged variables or derived variables like the sum of fatalities of the neighboring areas \citep{lindholm2022predicting}. However, forecasting conflict remains challenging, especially at finer geographical scales, because conflict data is highly skewed, often with zero values. Despite improvements, most models still face performance issues due to limited flexibility. A recent conflict forecasting competition by the Views team \citep{vesco2022united} concluded that more flexible models tend to perform better. While newer approaches are reducing model rigidity, such as the non-parametric smoothing method used by \cite{racek2024integrating}, they remain insufficient for fully capturing conflict dynamics. Models such as log-Gaussian Cox processes have been adapted for conflict studies \citep{zammit2012point}, but their use remains limited: they focus primarily on the Afghan conflict and rely on WikiLeaks data, which is difficult to reproduce for global applications. 

In the last couple of years, deep learning models have gained popularity, like long short-term memory (LSTM) \citep{radford2022high,von2025next}, or even temporal fusion transformers \citep{walterskirchen2024taking}. Additionally, large language models are adapted to spatio-temporal patterns mining in other fields \citep{li2024urbangpt,zhang2024large}, and offer promising performances. However, those remain poor in interpretability and lack traceability in the predictions. 

In recent years, conflict forecasting studies have largely moved beyond country-level analyses. Some studies focus on the first or second-order administrative regions \citep{aas2011all}, but remain sparse. Most of them that employ finer geographic scales rely on the Prio-Grid system \citep{tollefsen2012prio}, which aggregates data, such as fatalities, into 0.5x0.5 degree cells (about 55x55 km at the equator). This system provides a consistent structure that offers a balance between geographic detail and the limited precision of event location data. Since conflict data is typically extracted from news reports, the exact locations are frequently vague or missing. Although UCDP includes information on geographic uncertainty that could be used to filter out imprecise events, most studies retain as many observations as possible. Given that conflict events are relatively rare, the priority is typically maximizing the number of observations rather than improving spatial accuracy. The appropriate level of aggregation remains a point of debate in the conflict studies literature. \citep{cook2022race}. 

Despite promising results from new methods, predicting conflict at a fine geographic scale remains very difficult. The data is highly zero-inflated and skewed at the country level \citep{cederman2017predicting}, and this problem becomes worse at smaller scales, such as the Prio-Grid level. To illustrate how rare these events are, only 0.2\% of observations in our study are non-zero. This extreme imbalance makes it hard to gather enough useful information in the training data to build an accurate model. As a result, most prediction models produce vague, smoothed outputs, similar to the results of convolutional smoothing in image processing. 

\begin{figure}[!h]
\centering
\begin{subfigure}{0.45\textwidth}
    \centering
    \includegraphics[width=\textwidth]{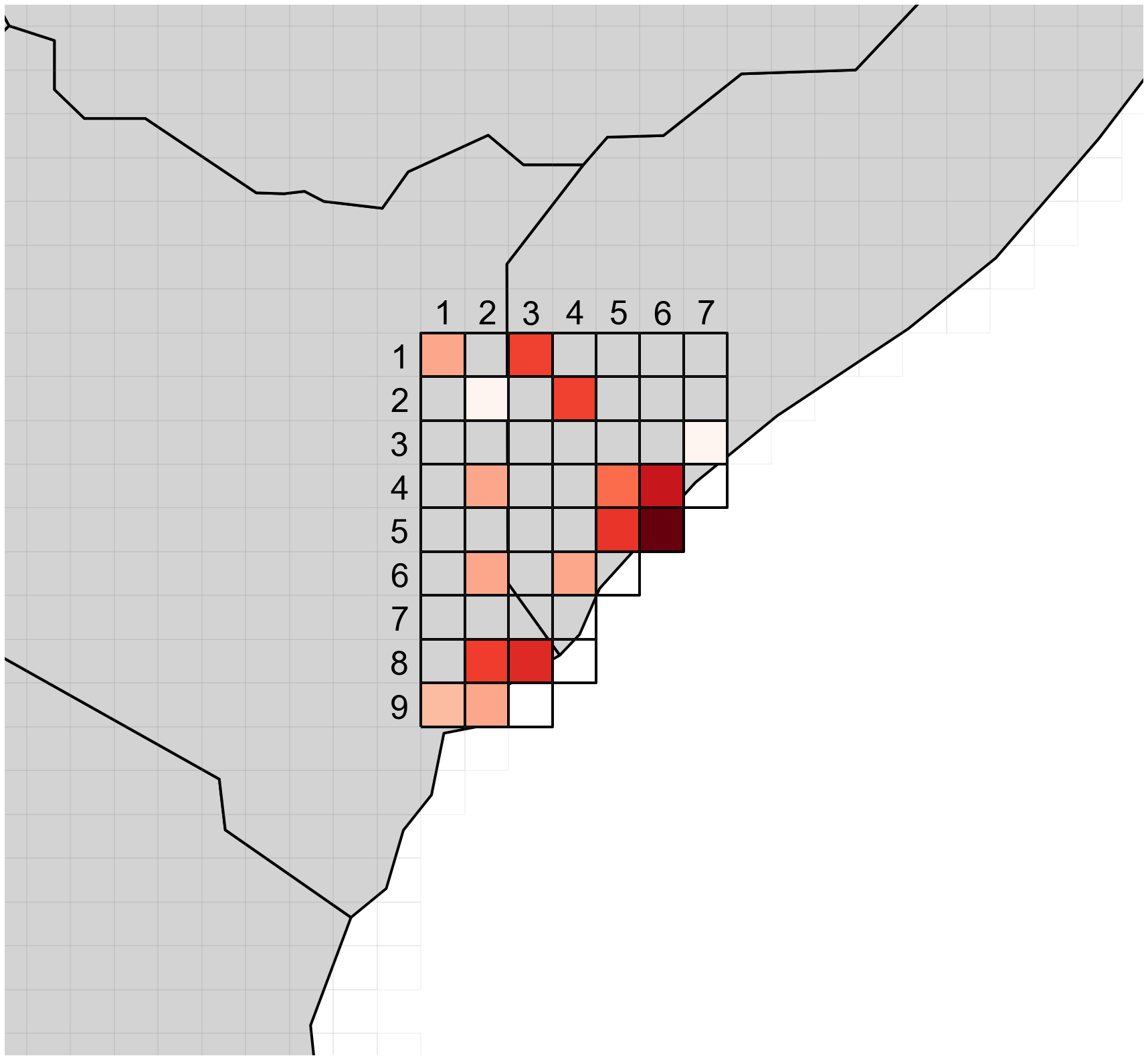}
    \caption{Input data - Cumulated fatalities from January 2022 to December 2022}
    \label{intro_why_1}
\end{subfigure}
\begin{subfigure}{0.45\textwidth}
    \centering
    \includegraphics[width=\textwidth]{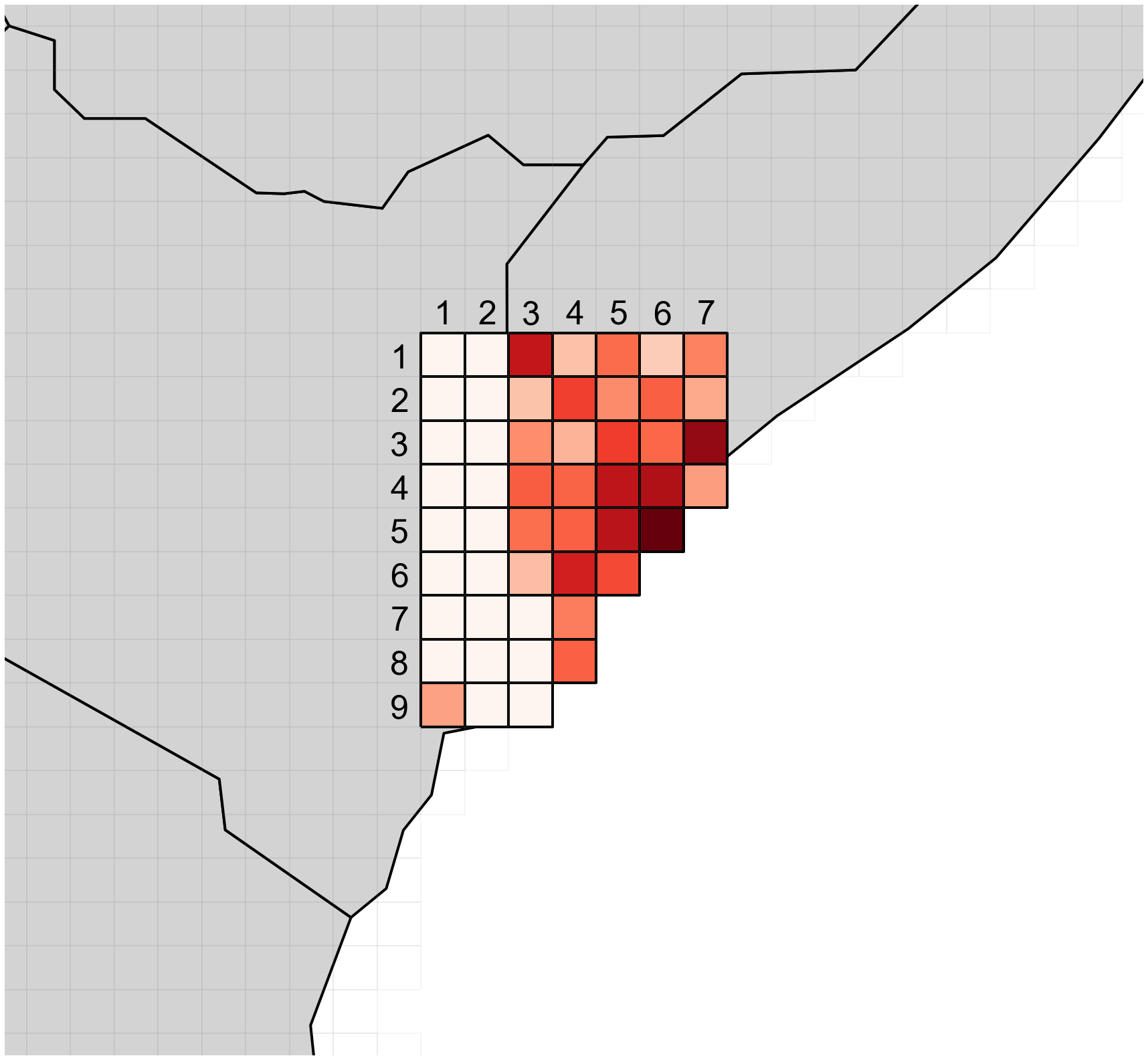}
    \caption{Views Forecast data - Predicted Cumulated fatalities from January 2023 to June 2023}
    \label{intro_why_2}
\end{subfigure}
\\
\begin{subfigure}{0.45\textwidth}
    \centering
    \includegraphics[width=\textwidth]{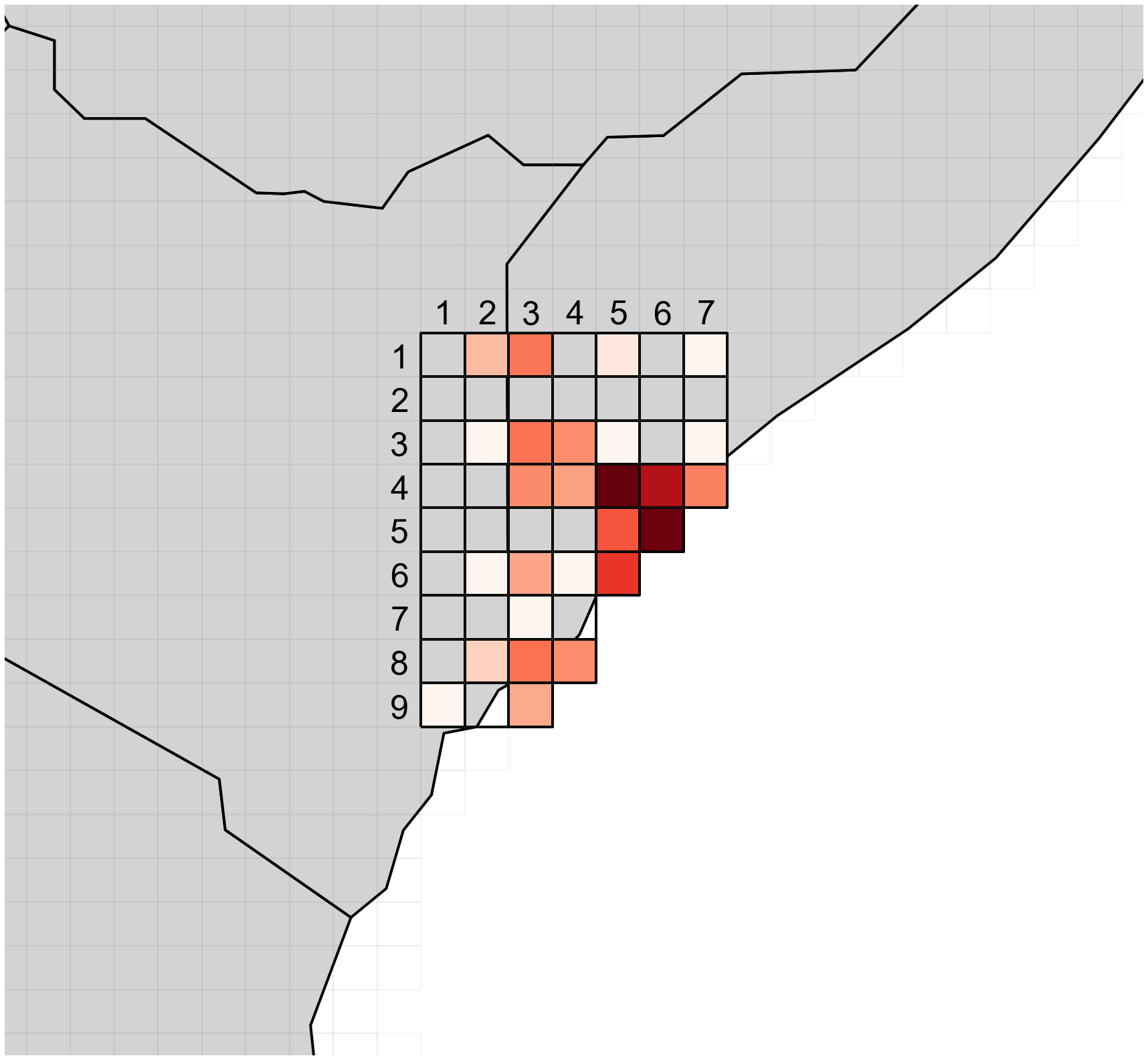}
    \caption{ShapeFinder Forecast data - Predicted Cumulated fatalities from January 2023 to June 2023}
    \label{intro_why_3}
\end{subfigure}
\begin{subfigure}{0.45\textwidth}
    \centering
    \includegraphics[width=\textwidth]{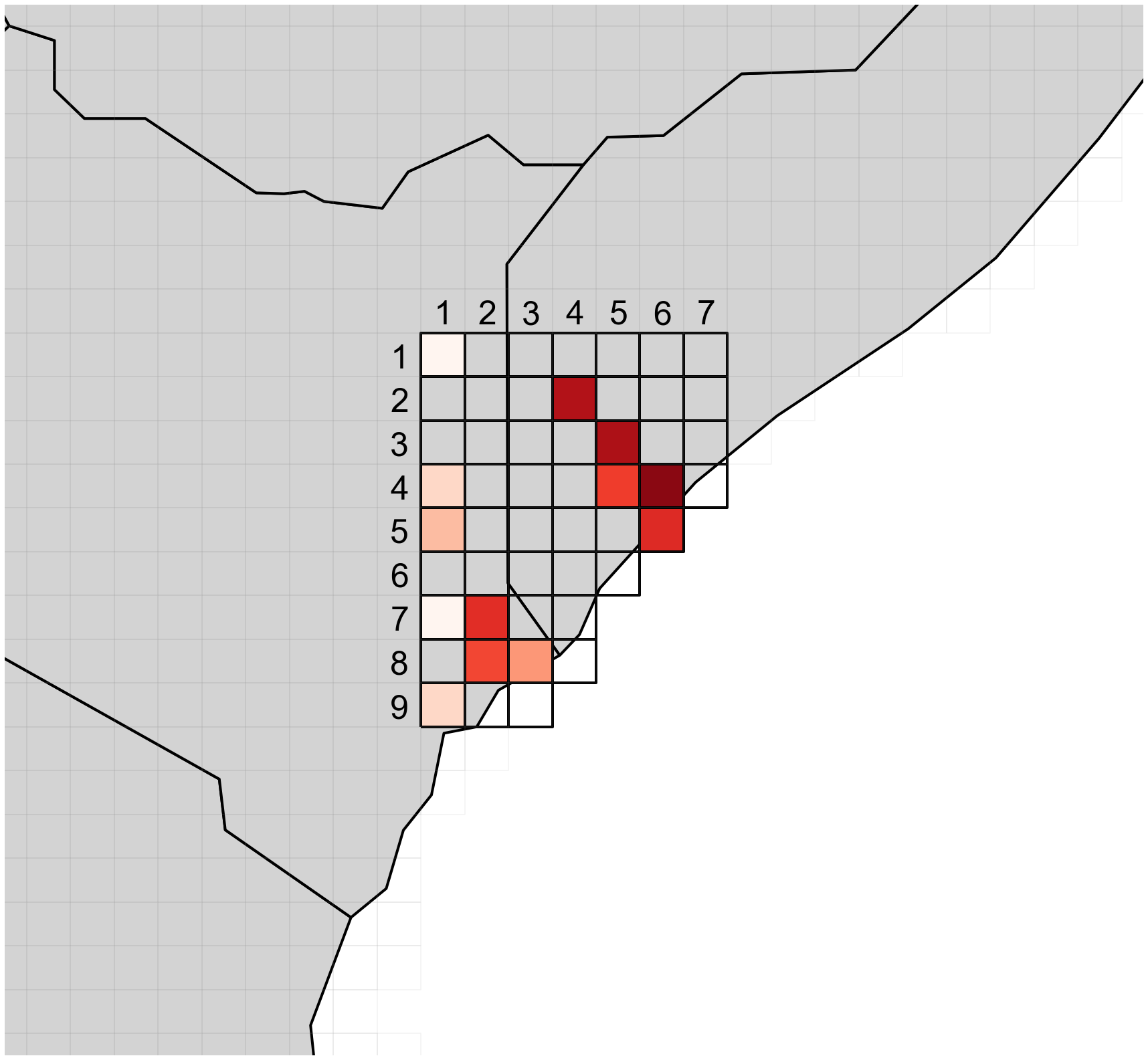}
    \caption{Observed data - Cumulated fatalities from January 2023 to June 2023}
    \label{intro_why_4}
\end{subfigure}
\caption{Conflict predictions for the Kenya–Somalia border during the first half of 2023, from Views (b), the ShapeFinder model (c) alongside the previous year’s cumulative fatalities used as model input (a) and the actual observed fatalities (d). Each colored square represents the cumulative fatalities in a PRIO grid cell—darker red indicates more fatalities; transparent cells had none. The numbers in the rows and columns are only included for reference and do not carry any meaning.}
\label{intro_why}
\end{figure}

Figure \ref{intro_why} shows conflict predictions for the Kenya–Somalia border during the first half of 2023, from the current best open-source benchmark, the Violence \& Impacts Early-Warning System (Views) (Fig. \ref{intro_why_2}).In the figure, we compare the forecast of the ShapeFinder (Fig. \ref{intro_why_3}) alongside the previous year’s cumulative fatalities used as model input (Fig. \ref{intro_why_1}) and the actual observed fatalities (Fig. \ref{intro_why_4}). Each colored square represents the cumulative fatalities in a PRIO grid cell—darker red indicates more fatalities; transparent cells had none. The Views forecast closely mirrors the input data, predicting high fatalities in the four cells in rows 4-5 and columns 5-6, and medium levels in surrounding cells, resembling the effect of a convolutional blur. This gives a general sense of where conflict might occur, but it results in clear overprediction. In this example, Views predicted 19 cells with more than seven fatalities, ShapeFinder predicted 8, and only 7 were actually observed.


\section{The Prio-Grid problem}
\label{prio_section}
The fine-scale geographic detail provided by the Prio-Grid scale demands greater flexibility compared to country-level models. A significant challenge lies in the key assumption of models with spatially lagged variables: the relationships between each Prio-Grid cell and its neighbors are uniform across the grid. This assumption often breaks down in practice. For example, a city like Cairo in Egypt spans several grid cells. If conflict arises within the city, fatalities in two or three neighboring cells might increase rapidly. However, assuming the same relationship between cells within Cairo as between cities like Yaoundé and Douala in Cameroon is difficult to justify. The dynamics of conflict propagation within a single urban area and between two distinct urban centers are not comparable. An effective model should account for these distinctions, treating Cairo as a single cohesive pole while recognizing Yaoundé and Douala as separate but interconnected nodes.

\begin{figure}[!h]
\centering
\includegraphics[width=\textwidth]{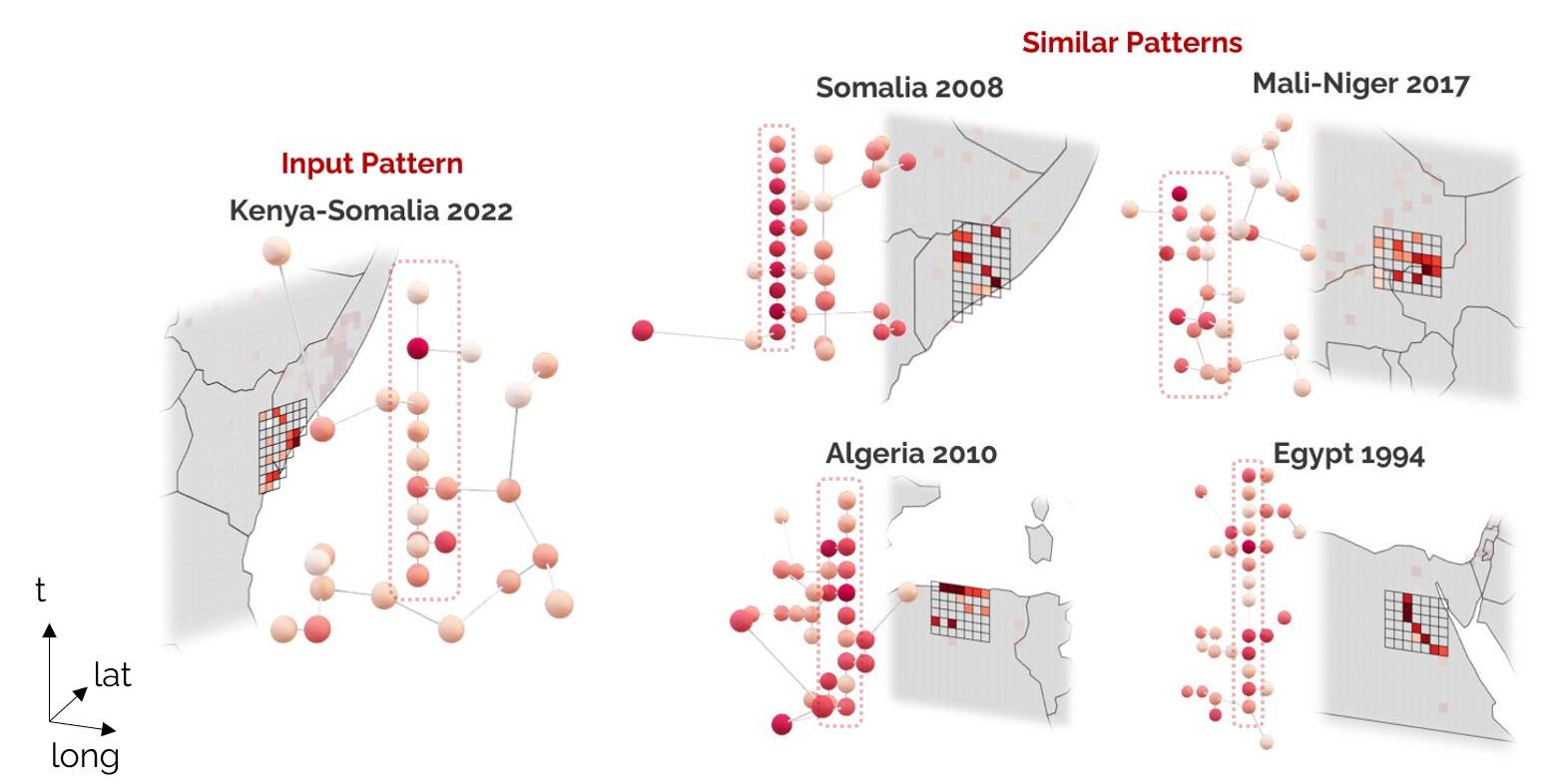}  \caption{3D sequences that happened at the border of Kenya and Somalia in 2022 (left section) and four similar cases found in the historical dataset by the model (right section). The 3D sequence uses longitude on the x-axis, latitude on the y-axis, and time on the z-axis. Each point represents the fatality value for a Prio-Grid cell in a specific month. Darker red points indicate higher fatality values, while gray links between points assist in visualizing the 3D structure but carry no interpretative meaning. Accompanying each 3D sequence is a map with accumulated fatalities over the period per Prio-Grid cell. The black grid represents the event’s location, and the colored squares indicate the total fatalities accumulated during the entire period within the Prio-Grid cells.}
\label{expla_test}
\end{figure}

Figure \ref{expla_test} shows an example of an input 3D sequence that happened at the border of Kenya and Somalia in 2022 (left section), corresponding to the three-dimensional representation of the example presented in Figure \ref{intro_why}. On the right section,  the four most similar cases found by the model within the historical dataset are displayed. The 3D sequence uses longitude on the x-axis, latitude on the y-axis, and time on the z-axis. Each point represents the fatality value for a Prio-Grid cell in a specific month. Darker red points indicate higher fatality values, while gray links between points assist in visualizing the 3D structure but carry no interpretative meaning. Accompanying each 3D sequence is a map with accumulated fatalities over the period per Prio-Grid cell. Similarly to Figure \ref{intro_why} format, the black grid represents the event’s location, and the colored squares indicate the total fatalities accumulated during the period within the Prio-Grid cells. 

Across the five patterns, a consistent structure emerges: one cell or a small cluster of cells accounts for most fatalities, maintaining steady intensity over the entire period (highlighted in dotted red on the figure). Occasionally, spreads of conflict appear in surrounding cells, but they rarely persist for more than a few months. Visually, these patterns resemble a tree, where the core cells with constant fatalities form the trunk, and the short-term spreads constitute branches.

Although the structures appear visually similar, detecting these patterns using a model is significantly challenging. One reason is the varying dimensions in latitude and longitude across the patterns: the Kenya-Somalia case is compact, while others, like Algeria or Mali-Niger, span multiple cells. Another complication is the difference in temporal durations: for example, the Egypt pattern extends over 16 months, whereas the Kenya-Somalia case lasts only 12 months. Additionally, the ``trunk'' structure varies considerably; in Somalia, it is concentrated in a single cell, while in Algeria, it spreads across two cells, and in Mali-Niger, it involves a group of four cells.

\begin{figure}[h]
\centering
\includegraphics[width=0.7\textwidth]{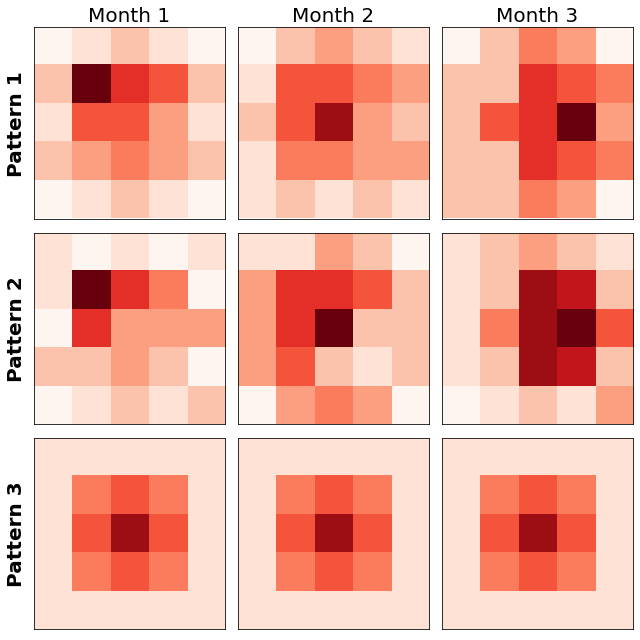}  
\caption{Three patterns on a 5×5 grid over three months. The x-axis represents longitude, and the y-axis represents latitude. The cell color indicates the number of fatalities per month, ranging from 0 (white) to 9 (dark red), with darker shades representing higher values.}
\label{toy}
\end{figure}

Traditional models, which rely on fixed spatial and temporal lags or neighbor sums, would fail to adequately identify the underlying similarity between these patterns. As a simplified toy model, Figure \ref{toy} presents three patterns on a 5×5 grid over three months. As in the previous figure, the x-axis represents longitude, and the y-axis represents latitude. The cell color indicates the number of fatalities per month, ranging from 0 (white) to 9 (dark red), with darker shades representing higher values. Visually, patterns 1 and 2 appear to have more similar dynamics, while pattern 3 is distinct, with constant values throughout the three months. However, when applying classical spatial regression methods to measure distances between patterns, the results indicate a smaller distance between patterns 1 and 3 than between patterns 1 and 2. In contrast, the flexible method introduced in this paper correctly identifies a smaller distance between patterns 1 and 2. Detailed calculations are provided in \ref{app:calcu}.

Incorporating flexibility into models is essential for matching cases with similar dynamics despite differences in speed, scale, or frequency. This flexibility has shown clear improvements even in single-dimensional models \citep{schincariol2025accounting}, but when extended to three dimensions, its importance becomes even more pronounced. Events can vary in velocity across three axes—space and time—introducing additional layers of complexity that rigid models cannot capture. 

\section{Data Creation}

\subsection*{Data source}

The data on conflict fatalities used in this study come from the Uppsala Conflict Data Program (UCDP) Georeferenced Event Dataset (GED) \citep{sundberg2013introducing,davies2022organized}. This dataset provides monthly data at the Prio-Grid level, covering the period from 1989 to the present. To validate our forecasting results, we compared them with the Views forecast ensemble model \citep{hegre_forecasting_2022}, which is a leading benchmark in conflict prediction. The Views model is an ensemble model that includes various machine learning models, including a large number of covariates. Since the model only forecasts the state-based fatalities in the Middle East and Africa, our study focuses on these regions and variable to enable a direct comparison. The data was obtained through the Views API.\footnote{The datasets $fatalities001\_2022\_00\_t01$, $fatalities001\_2022\_06\_t01$, $fatalities001\_2023\_00\_t01$, $fatalities001\_2023\_06\_t01$, produced by Views are available here: \href{https://github.com/prio-data/views_api/wiki/Available-datasets}{https://github.com/prio-data/views\_api/wiki/Available-datasets}}

\subsection*{Active zones identification}

\begin{figure}[!h]
\centering
\begin{subfigure}{0.45\textwidth}
    \centering
    \includegraphics[width=\textwidth]{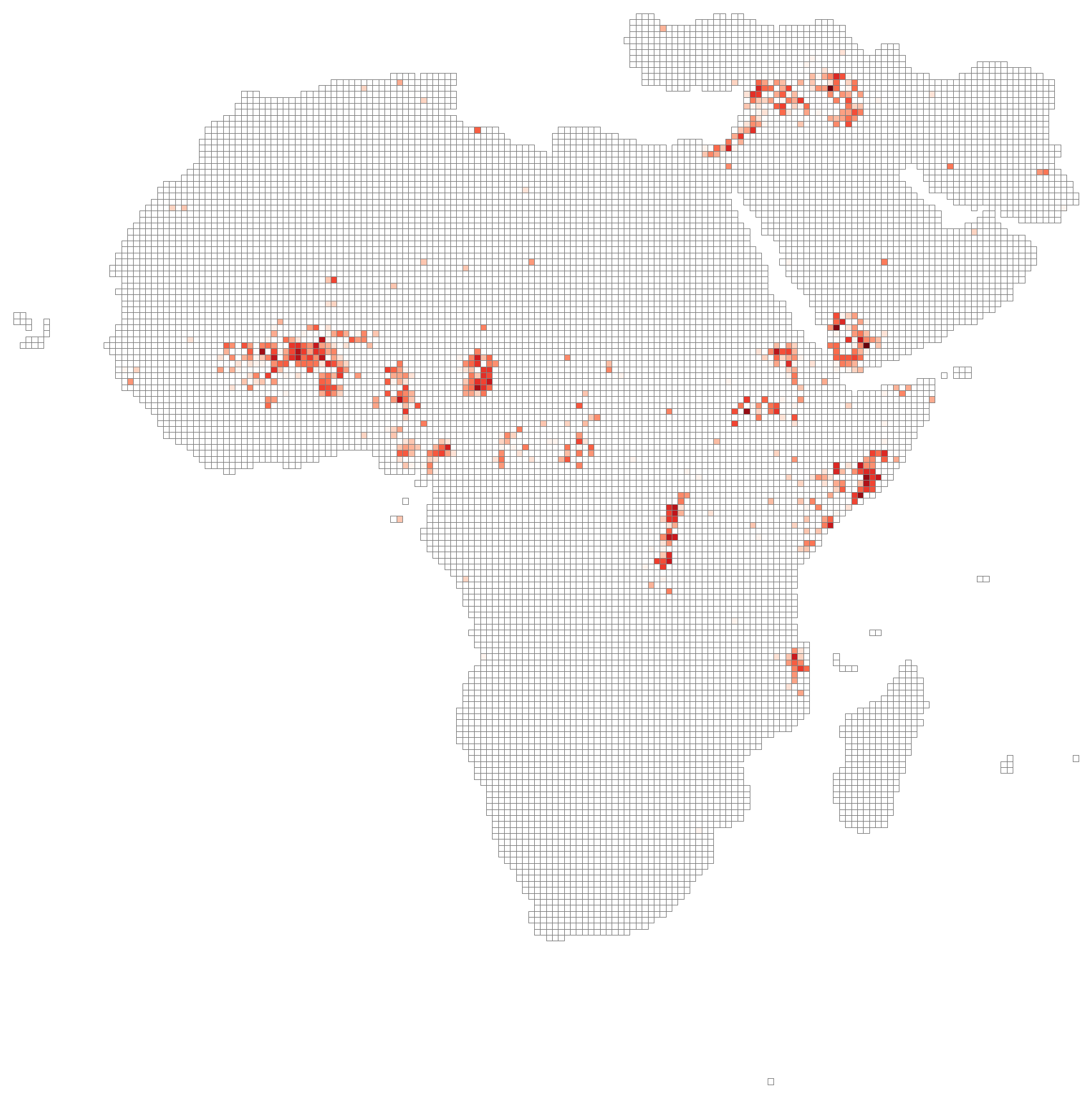}
    \caption{Cumulated fatalities from January 2022 to December 2022}
    \label{ex_pattern_1}
\end{subfigure}
\begin{subfigure}{0.45\textwidth}
    \centering
    \includegraphics[width=\textwidth]{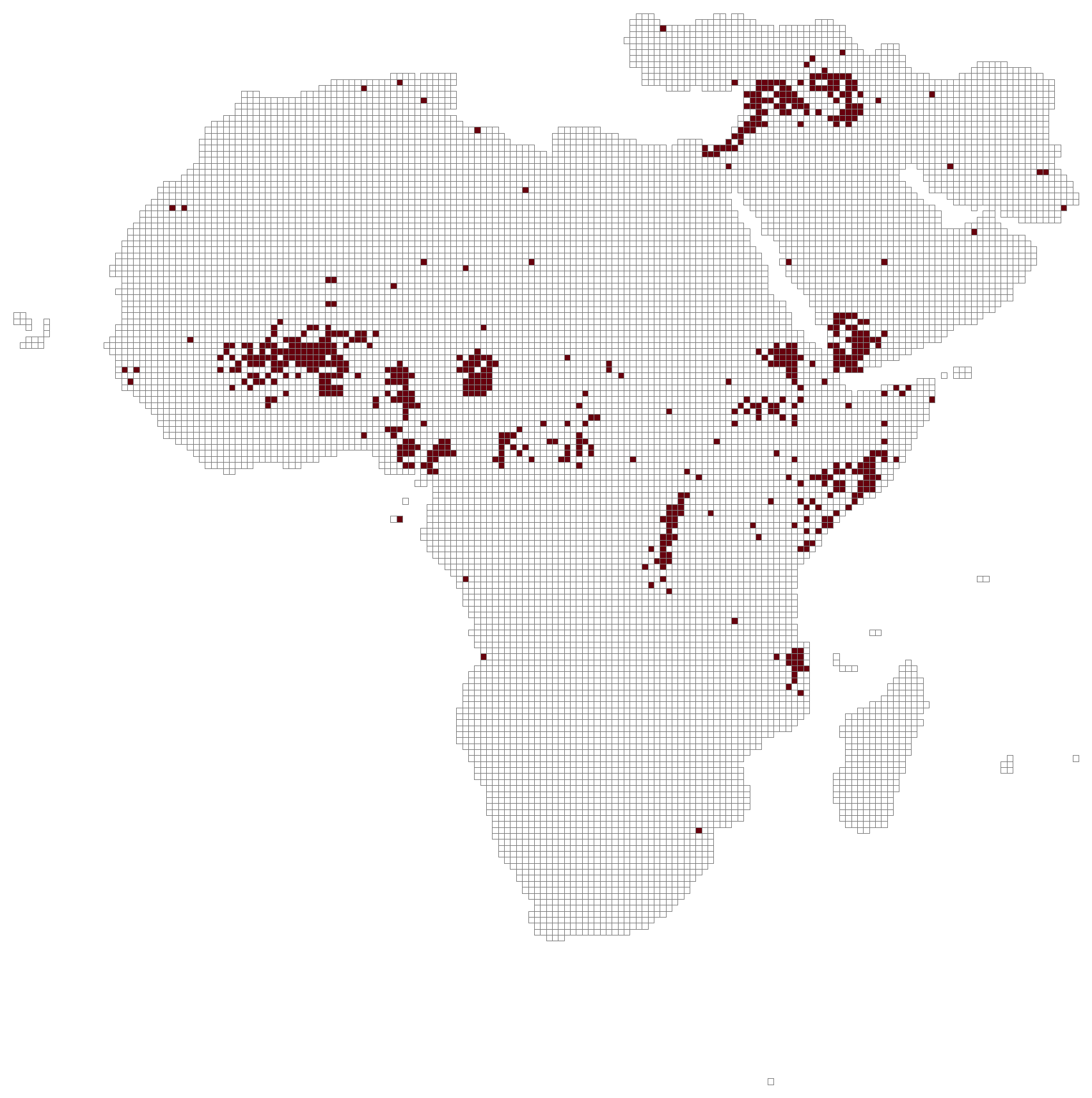}
    \caption{Binary classification of active cell, with at least one cumulated conflict fatalities}
    \label{ex_pattern_2}
\end{subfigure}
\\
\begin{subfigure}{0.45\textwidth}
    \centering
    \includegraphics[width=\textwidth]{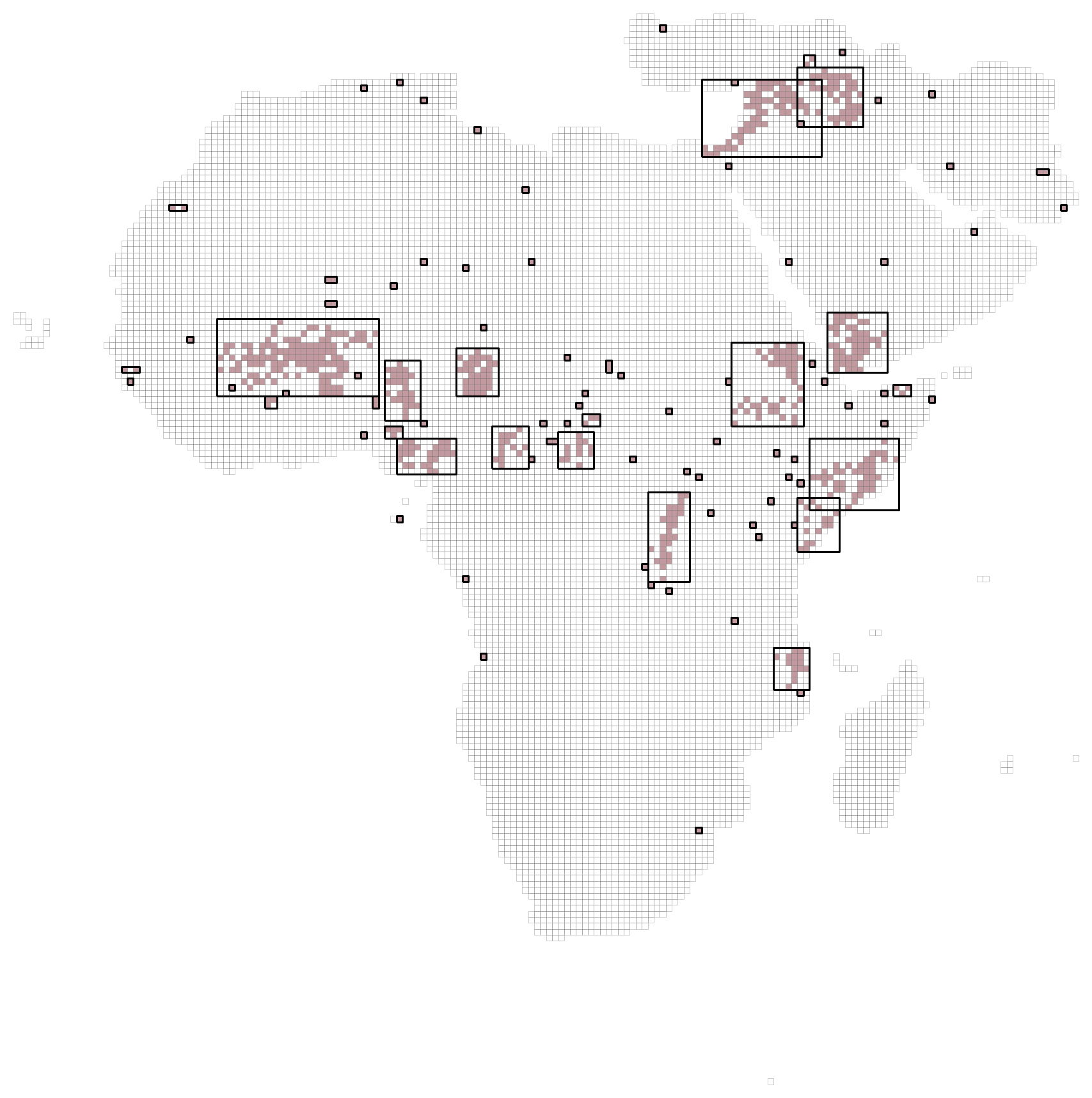}
    \caption{Outcome of the Connected Component Labeling. The black borders outline the identified patterns.}
    \label{ex_pattern_3}
\end{subfigure}
\begin{subfigure}{0.45\textwidth}
    \centering
    \includegraphics[width=\textwidth]{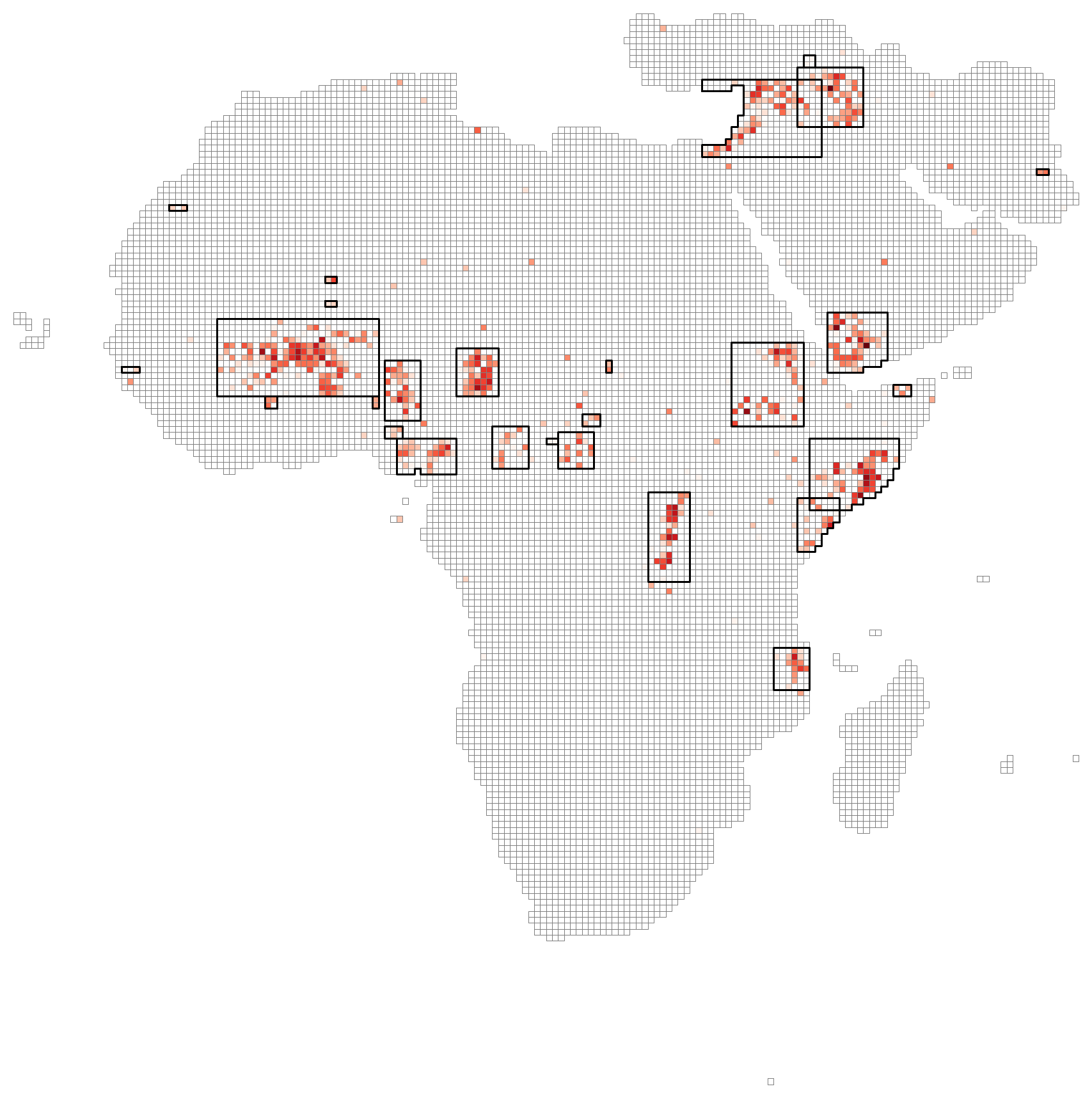}
    \caption{Union-find patterns without filtered isolated cells by morphological erosion}
    \label{ex_pattern_4}
\end{subfigure}
\caption{Four steps of the identification of active conflict zone patterns in 2022 in Africa and the Middle East. The red shading represents the intensity of conflict, with darker red indicating higher numbers of fatalities in (a) and (d).}
\label{ex_pattern}
\end{figure}

The first step of our model is to identify potential future zones of conflict by applying object-detection techniques from image analysis, like connected-component labeling and morphological erosion. The four steps of the method are represented in Figure \ref{ex_pattern}. This process begins by examining the previous 12 months prior to the forecast period. Fatalities from the previous year are aggregated within each Prio-Grid cell, resulting in a 2D grid where each cell contains the sum of fatalities for the year (Fig. \ref{ex_pattern_1}). The model then scans the grid using a rolling window across latitude and longitude to identify cells with non-zero values, referred to as ``active'' cells (Fig. \ref{ex_pattern_2}).

Once the active cells are identified, Connected Component Labeling (CCL), a well-established technique in image analysis, is applied (Fig. \ref{ex_pattern_3}). CCL is commonly used in image analysis to detect and classify objects based on pixel connectivity. Once an active cell is identified, its neighboring cells within a 2-unit radius are examined. If a neighboring cell is also active, it is grouped into the same cluster. This iterative process continues until no additional active neighboring cells can be found. The resulting interconnected active cells are classified as `active conflict zone'. The radius is set to 2 units, corresponding to approximately 100km, allowing flexibility of the detection and remaining relatively close enough to have a link. 
The final step applies a modified version of morphological erosion and dilation (Fig. \ref{ex_pattern_4}). Erosion removes small structures by eroding object boundaries, eliminating minor, irrelevant details. Dilation expands objects, fills small gaps, and reconnects structures. These techniques are commonly used in remote sensing, such as for cloud detection, where erosion removes noisy pixels that may be mistaken for small clouds, and dilation merges fragmented cloud pixels into a coherent shape. Here, the erosion is applied by retaining only zones with more than one active cell. Indeed, as shown in Figure \ref{app:projo}, isolated active cells are associated with future conflict only 13\% of the time and never diffuse to the neighboring cells. A spatial model would not be appropriate. However, these excluded cells may present opportunities for future work, such as applying a non-spatial autoregressive model as described in \cite{schincariol2025accounting}, though their exclusion does not significantly impact the overall evaluation (see the Discussion section). The dilation is done by the union of entirely overlapping identified zones. If two or more active zones share some cells, the mean of the predictions is considered. 

Using this active zone identification method, 99\% of the total fatalities and 94\% of active cells are considered. The remaining 6\% of active cells are isolated and not linked to any larger zones. This aggregation method provides a compromise between maximizing the inclusion of relevant fatalities for effective model training and maintaining computational efficiency. By doing so, the model remains capable of running in just a few hours. Further aspects of the impact of this selection are explored in \ref{imp_acs}.

\subsection*{Data transformation}\label{data_trans}

\begin{figure}[h]
\centering
\includegraphics[width=\textwidth]{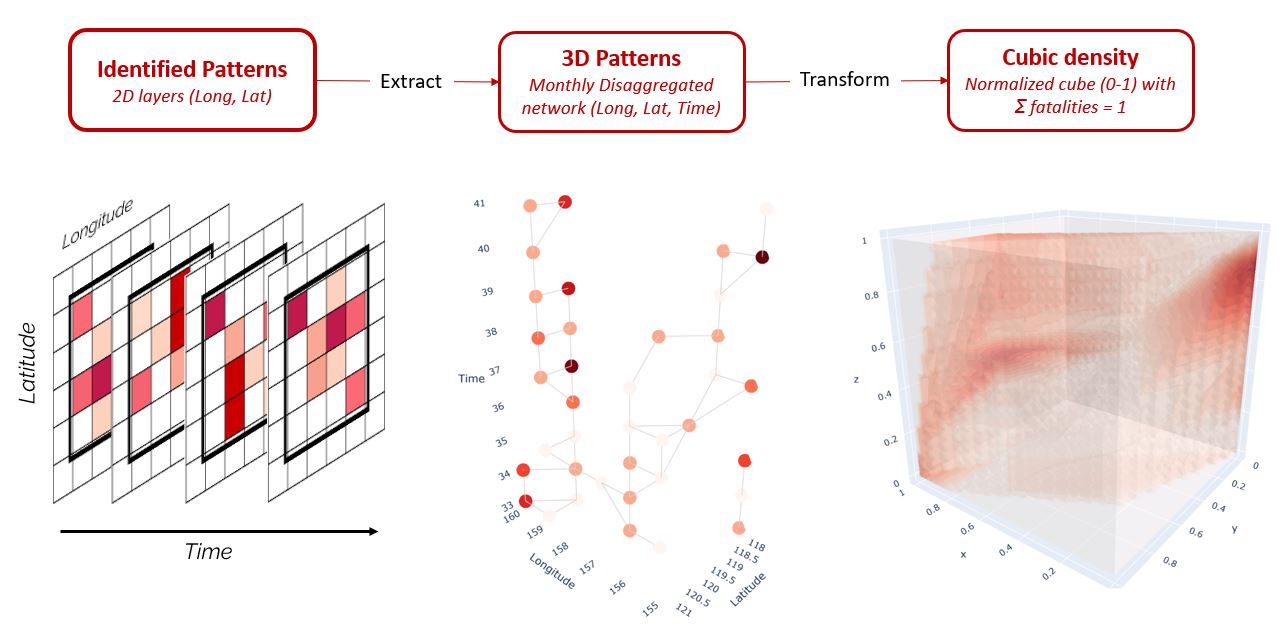}  \caption{Data transformation process illustrated through three steps. (1-left) Active zones are identified in the previous 12-month aggregated 2D grid of active cells (2-middle) 3D sequences of linked neighboring cells with conflict fatalities are extracted and represented in a ``molecular'' format, where Longitude is on the x-axis, Latitude on the y-axis, and Time on the z-axis. Darker red shades indicate higher fatality values. Gray links between cells aid in 3D visualization but do not signify any meaningful connections. (3-right) The extracted sequences are transformed into density cubes, with each side representing a single normalized value and the intensity of conflict as the density value. As in (2), darker red indicates higher density values.}
\label{trans}
\end{figure}

To implement our shape-matching method, the raw data undergoes a transformation process, as illustrated in Figure \ref{trans}. After the active zones identification, we extract groups of conflict cells from the accumulated layers of 2D grids. The 2D zones are then disaggregated into a 3D sequences with Longitude and Latitude as the x-axis and y-axis, and month as the z-axis. A sequence example is shown in the middle part of the figure, where each colored point represents a Prio-Grid/month cell, with darker red indicating higher fatalities. The gray lines connecting the points are added for visual clarity and do not carry any specific meaning. In this example, the points are connected, forming a sequence with two branches—one spanning longitudes 160 to 159 and the other from 157 to 155—both connected by a common base around time 33 to 35.

Finally, the sequence is transformed into a density cube, as shown on the right side of the figure. The three axes: latitude, longitude, and time, are normalized from 0 to 1, and this spatial and temporal normalization includes non-active cells (those without fatalities) within the active zone. This spatial and time normalization allow the comparison of 3D sequences with different geographic and time dimensions. The total number of fatalities within the cells is also normalized so that their sum equals 1, similar to a standard density function, to apply the Earth Moving Distance (EMD) to measure the difference in shape between two sequences. This final step results in a cube with 1 unit per side, containing a 3D density distribution.

\section{Model Methodology}

\begin{figure}[!h]
\centering
\includegraphics[width=0.85\textwidth]{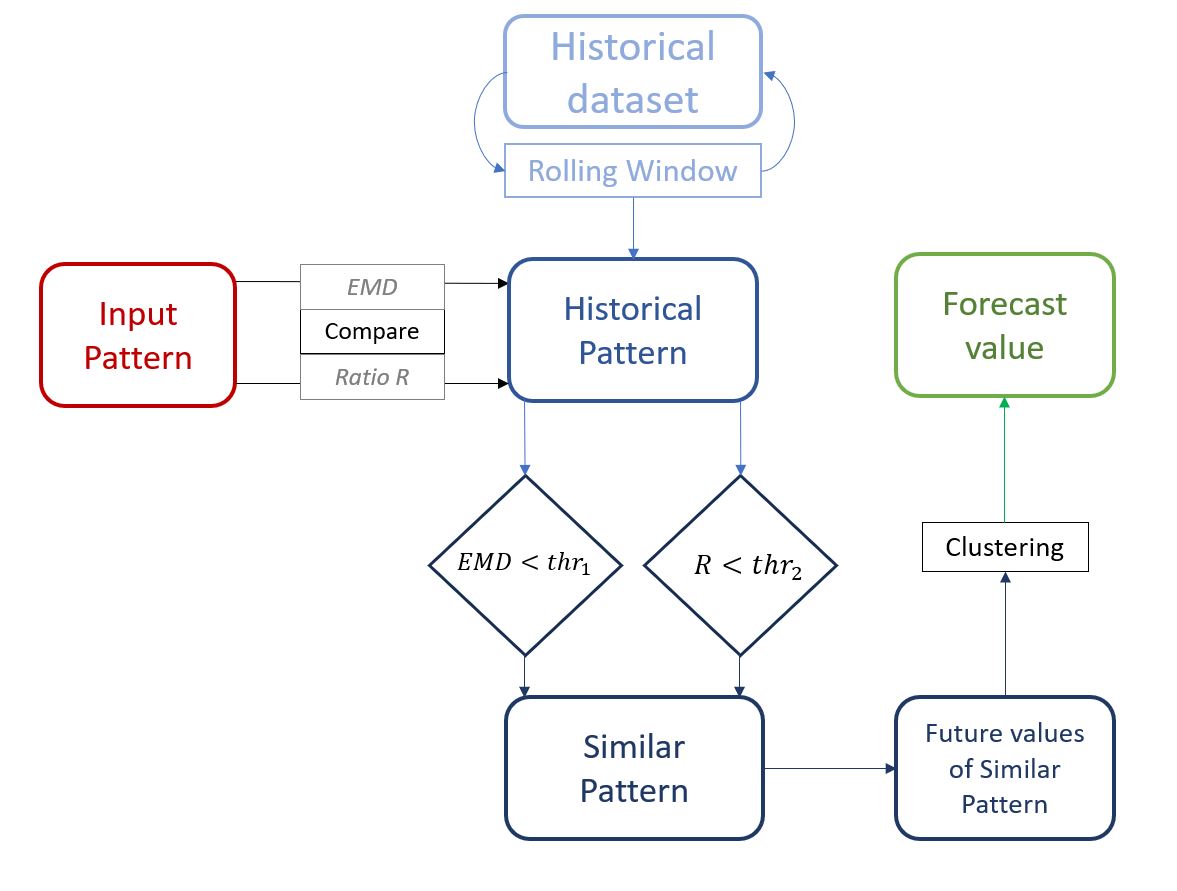}  \caption{Diagram of the overall method of the model. First, we extracted the 3D sequence of interest as mentioned before, called input sequence in the figure. Then, we look for similar cases in the historical data using a rolling window, by calculating two distance metrics, Earth Moving Distance (EMD) and ratio of active cells (R). If those two metrics are lower than a threshold, the historical sequence is classified as a similar case to the input one. Finally, the values that follow similar cases are clustered to produce the forecast values.}
\label{recap}
\end{figure}

The structure of the model is illustrated in Figure \ref{recap}. The process begins with the extraction of sequence of interest (see Data transformation), referred to as input sequence in the figure. Next, the model searches for similar cases in the historical data using a rolling window. This involves calculating two distance metrics: the Earth Mover’s Distance (EMD) and the ratio of active cells (R). If both metrics fall below a defined threshold, the historical sequence is classified as similar to the input. Finally, the subsequent values associated with these similar cases are clustered to generate the forecast values. The definitions of key concept during the several steps of the modeling are included in \ref{app:def}.

\subsection*{Distance between two 3D Pattern}

The distance metrics for two 3D patterns combine two comparison factors: (1) the Earth Moving Distance (EMD) between the cubic density representations of the patterns, and (2) the ratio of cells with at least one conflict between the two patterns. 

The Earth Mover’s Distance (EMD) \citep{andoni2008earth} quantifies the similarity between two distributions. Initially developed for transportation logistics, it has been adapted for applications in image processing and computer vision. The metric evaluates the amount of work needed to transform one distribution into another. This is achieved by solving an optimal transport problem, minimizing the total cost required to transport the mass. The EMD is defined by the following equation:

\begin{equation}
    \text{EMD}(P, Q) = \min_{\gamma \in \Gamma(P, Q)} \sum_{i,j} \gamma_{ij} \, c(x_i, y_j)
\end{equation}
\text{where:}
\begin{itemize}
    \item \( P \) and \( Q \) are the two distributions of points with non-null values for each pattern.
    \item \( \Gamma(P, Q) \) is the set of all possible joint distributions (couplings) with marginals \( P \) and \( Q \).
    \item \( \gamma_{ij} \) represents the difference in normalized fatalities from \( x_i \) to \( y_j \).
    \item \( c(x_i, y_j) \) is the Euclidean distance between \( x_i \) to \( y_j \).
\end{itemize}

It is important to note that we compare each sequence not once but four times. The cubic density is rotated 90° around the time axis three times more because a sequence that starts in the southeast and ends in the northwest could be similar to one that starts in the southwest and ends in the northeast. The rotation is done around the time axis that remains fixed, as time progresses in only one direction. We take the minimum EMD value of the four sequence orientations. 
This distance metric allows us to identify similar sequences that may vary in duration and geographic spread—patterns that traditional models often fail to capture. By accommodating variations in both time and space, the method provides a more adaptive understanding of conflict dynamics, enabling the detection of similar cases that may be overlooked by classical models due to their rigidity. This flexibility is crucial for accurately analyzing complex phenomena where the timing and geographic extent of events can differ significantly. 

The second part of the ratio of active cells (R) of the two sequences. An active cell is a Prio-Grid month observation that has a non-zero value. The R ratio is defined by the following equation:

\begin{equation}
    \text{R} = \left|\tanh\left(\log\left(\frac{N_{p1}}{N_{p2}}\right)\right)\right|
\label{eq2}
\end{equation}
\text{where:}
\begin{itemize}
    \item \ $N_{p1}$, $N_{p2}$ is the number of conflict active cells in the first and second sequence
    \item \ $R$ is the Ratio of Active Cells
    \item \ $tanh$ is the hyperbolic tangent function
\end{itemize}

EMD alone does not account for the counts of conflicts or whether the sequences have a similar number of active cells, those with non-zero fatalities. Thus, two sequences could have similar shapes (low EMD) but very different counts of active cells. Although the model assumes that sequences with different magnitudes are comparable, it does not equate cells with zero fatalities to those with a small number of fatalities. The Ratio of Active Cells function \ref{eq2} is equal to zero when the ratio $\frac{N_{p1}}{N_{p2}}$ is equal to 1. It increases in the same way whether the ratio is negative or positive, which removes the need to control the numerator and denominator separately.

The first factor ensures that the overall shapes of the patterns are similar, while the second ensures that both patterns have a comparable number of active conflict cells.
By combining both factors, this method provides a more comprehensive and reliable approach to pattern matching. Two patterns are considered similar if they fall below two critical threshold values.

\subsection*{Search for similar patterns in the historical data}

\begin{figure}[h]
\centering
\includegraphics[width=0.8\textwidth]{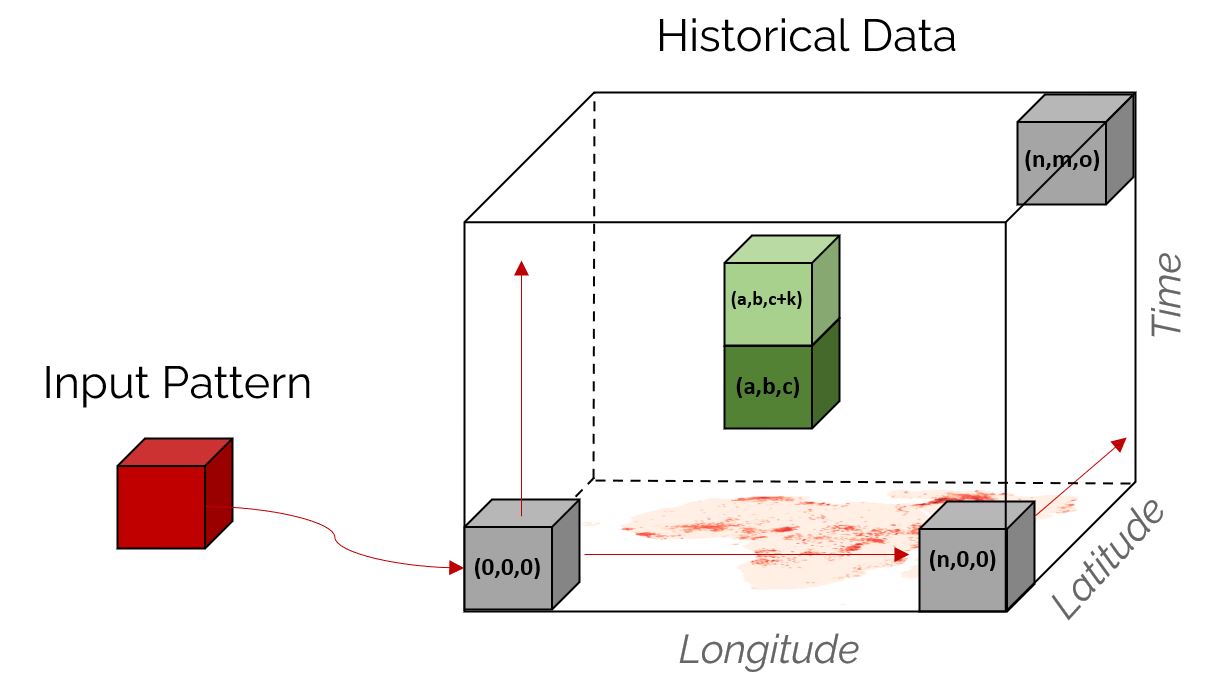}  \caption{Rolling Window Scheme for historical search. The red cube represents the input sequence, while the gray cubes show the historical dataset, covering a range from 0 to n in longitude, 0 to m in latitude, and 0 to o in time. The dark green cube represents a similar case to the input, with the light green cube showing its following values.}
\label{hist_search}
\end{figure}

To identify similar historical cases for a given input sequence, we use a rolling window method, as detailed in \ref{hist_search}. Input sequences (highlighted in red in the figure) are compared to historical sequences spanning from southwestern Africa in 1989 to the northeastern Middle East in the final month of the training set (for example, December 2021 for the first test period, which forecasts January to June 2022). The rolling window moves sequentially through longitude, latitude, and time.

For instance, the first historical sequence to be compared covers 0 longitude to $i$, 0 latitude to $j$, and 0 time to $k$, where $i,j,k$ are the dimensions of the input. The next sequence shifts by half the size of the longitude dimension, from $(i/2)$ longitude to $i+(i/2)$, while keeping latitude and time fixed. This process repeats similarly through the latitude and time axes. By shifting the window by half the dimension size, we minimize overlap between patterns and shorten computation time.

For each iteration of the rolling window, sequences with similar dimensions (longitude, latitude, and time) are compared, but also sequences of ±1/4 in each dimension. If a historical sequence meets the similarity criteria, it is marked as similar (dark green in the figure), and its subsequent values, referred to as ``past future'' throughout the paper (light green in the figure), are stored. The outcome of the rolling window is a set of similar sequences and their following values.

\subsection*{Forecast creation}

The forecast values are generated using the ``past future'' of similar historical sequences. This process has two main steps: (1) reshaping the past future values to fit the input dimensions, and (2) clustering and identifying the most likely scenario. Because we allow comparing sequences with different dimensions, some similar sequences of the same pattern may differ in size. Extending a sequence is done by multiplying dimensions, while larger sequences are shrunk by dividing dimensions to fit the input size. Forecast values are then adjusted to align with the closest matching grid cells. If multiple values fall in the same cell, they are added together. 

\begin{figure}[!h]
\centering
\includegraphics[width=0.8\textwidth]{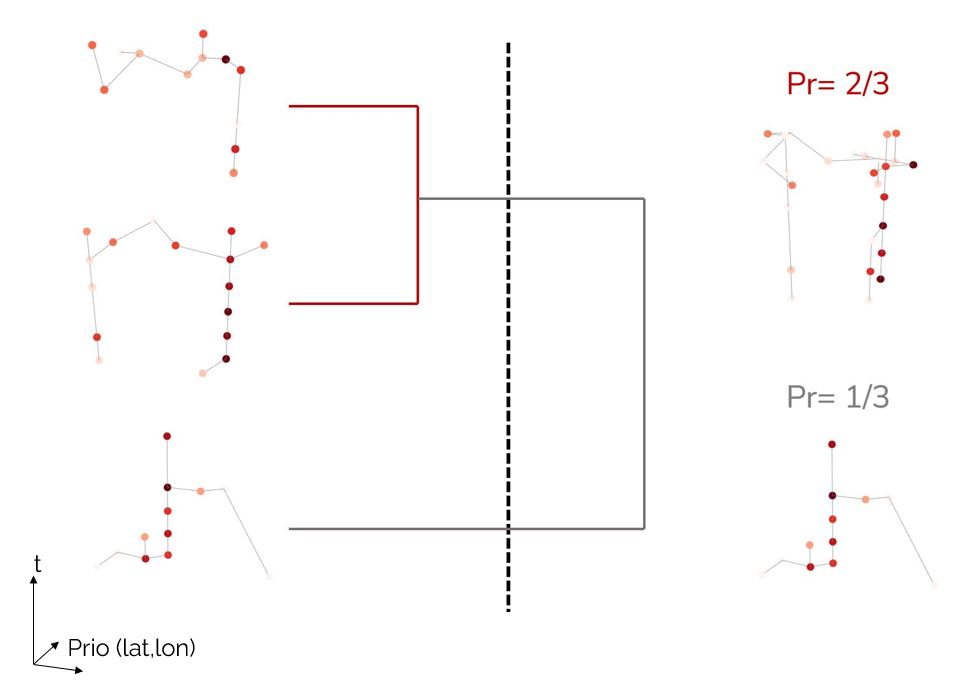}
\caption{Example of scenario creation using clustering.  The 3D sequence uses longitude on the x-axis, latitude on the y-axis, and time on the z-axis. Each point represents the fatality value for a Prio-Grid cell in a specific month. Darker red points indicate higher fatality values, while gray links between points assist in visualizing the 3D structure but carry no interpretative meaning. Three similar patterns were identified in the historical data (left) and created two scenarios (right) with their associated probability, based on the number of observations in each cluster.}
\label{clus_ex}
\end{figure}

Next, we apply clustering on past futures to find the most probable forecast scenario. After mapping each sequence’s projection to longitude, latitude, and time, we calculate the Euclidean distance between these projections and cluster them based on a set threshold distance. Similarly to the previous distance parameters, the clustering threshold was established based on a parameter grid search to minimize the evaluation metrics of the validation set. Scenarios are then created by grouping the values by cluster using the mean value for each cell point (longitude, latitude, month). The number of observations per cluster estimates the probability of observing this outcome emerge when this pattern appears. The forecasted value is then the mean value of the cluster with the most observations, representing the most probable future scenario. Figure \ref{clus_ex} presents an example of scenario creation using clustering. Three similar sequences were identified in the historical data (left). The two upper patterns have a distance below the defined threshold (indicated by the dotted line), and are then grouped together to form a scenario with a probability p=2/3. As this is the most probable scenario, it is selected as the forecast value. This step helps mitigate the influence of outliers, which could heavily weight the prediction with extreme values that rarely have occurred in history. 

\section{Evaluation}

\subsection*{Benchmark model}

We evaluate our model, labeled as `ShapeFinder'(SF), by comparing it to the leading benchmark in the field, the Views forecast ensemble model \citep{hegre_forecasting_2022}. This model is one of the few that provides a publicly available monthly point forecast at the grid level. It forecasts up to 36 months ahead for the Middle East and Africa, starting from January 2022.\footnote{It is important to note that their model is optimized with 36-month horizons, and these predicted values are actually live predictions.} Our evaluation is conducted across four 6-month periods:
\begin{itemize}
    \item January to June 2022, using data up to December 2021.
    \item July to December 2022, using data up to June 2022.
    \item January to June 2023, using data up to December 2022.
    \item July to December 2023, using data up to June 2023.
\end{itemize}

These intervals provide a balance between a long enough forecast horizon and sufficient observations. A six-month horizon, half the length of the 12-month input window, offers the best trade-off between dynamic prediction reliability and the potential operational utility. A shorter, three-month horizon may not be sufficient for a warning system, while a 12-month horizon likely exceeds the theoretically temporal limits of reliable dynamic prediction.

\subsection*{Metrics}

The models' performance is measured using three metrics. The first is the Mean Squared Error (MSE) at the Prio-Grid level, the most common metric in forecasting. MSE measures the average prediction error but does not account for flexibility in geography or time. It simply checks if the forecast at time t for Prio-Grid p is accurate. The formula for MSE is:
\begin{equation}
    MSE=\frac{1}{n}\sum_{i=1}^n(Y_i-\hat{Y_i})^2
\end{equation}
With $Y_i$ the observed value and $\hat{Y_i}$ the predicted value.

However, because Prio-Grid is a very fine-grained geographic scale, this metric alone can be misleading. For example, a forecast of 1,000 fatalities in cell c1 when the actual number is 1,000 in the neighboring cell c2 results in a large penalty (an error of +1,000 in c1 and –1,000 in c2), even though the prediction is only off by 55 km. For policymakers, such a small spatial imprecision is still highly valuable. To address this issue, two additional zone-level (defined in Fig.\ref{ex_pattern}) metrics were introduced. These capture the total number of fatalities within the zone over the forecast period, following the approach used by the ConflictForecast team \citep{mueller2024introducing}, which evaluates 3-month and 12-month cumulative fatalities. These metrics provide a clearer overview of forecast quality without penalizing the model for small temporal or spatial shifts. Potential users are typically more interested in whether fatalities will occur in the zone over the next six months than in perfect monthly alignment. To complement these measures and assess the predicted dynamics, the EMD is also used as an evaluation metric. Combined, these three metrics, active zone accumulated fatalities error, EMD for assessing dynamic similarity, and Prio-Grid MSE for fine-grained accuracy, offer a comprehensive evaluation of both overall trends and detailed spatial performance.

To extend the analysis from forecasting evaluation, we need to characterize a sequence and its relationship with its future values. For this, we introduce a metric called ratio of increase ($R_{inc}$), which quantifies the difference in the number of fatalities between a sequence and its past future. This metric helps assess whether a pattern leads to an escalation, stabilization, or decline in conflict intensity. The formula for $R_{inc}$ is:
\begin{equation}
    R_{inc}= \frac{\sum fat_{pf}}{\sum fat_{sequence}}
\end{equation}
With $\sum fat_{pf}$ the sum of fatalities in the past future section, and $\sum fat_{sequence}$ the sum of fatalities in the pattern.

\subsection*{Parameters optimization}

The three parameters of the model, the EMD threshold, the R threshold $thr_{2}$, and the cluster distance $clu_{d}$ are selected with a grid-search on a validation set from January to June 2021, using data up to December 2020 in the training set, minimizing the MSE at the Prio-Grid level. The parameters that lead to the best performance are $thr_{1} = 0.15$ and $thr_{2} = 0.05$, and cluster distance $clu_{d} = 0.0054*N_{dim}$ where $N_{dim}$ is the number of Prio-Grid monthly elements in the forecasted sequence.

\section{Results}

\subsection*{Patterns and Past Futures}

\begin{figure}[!h]
\centering
\includegraphics[width=\textwidth]{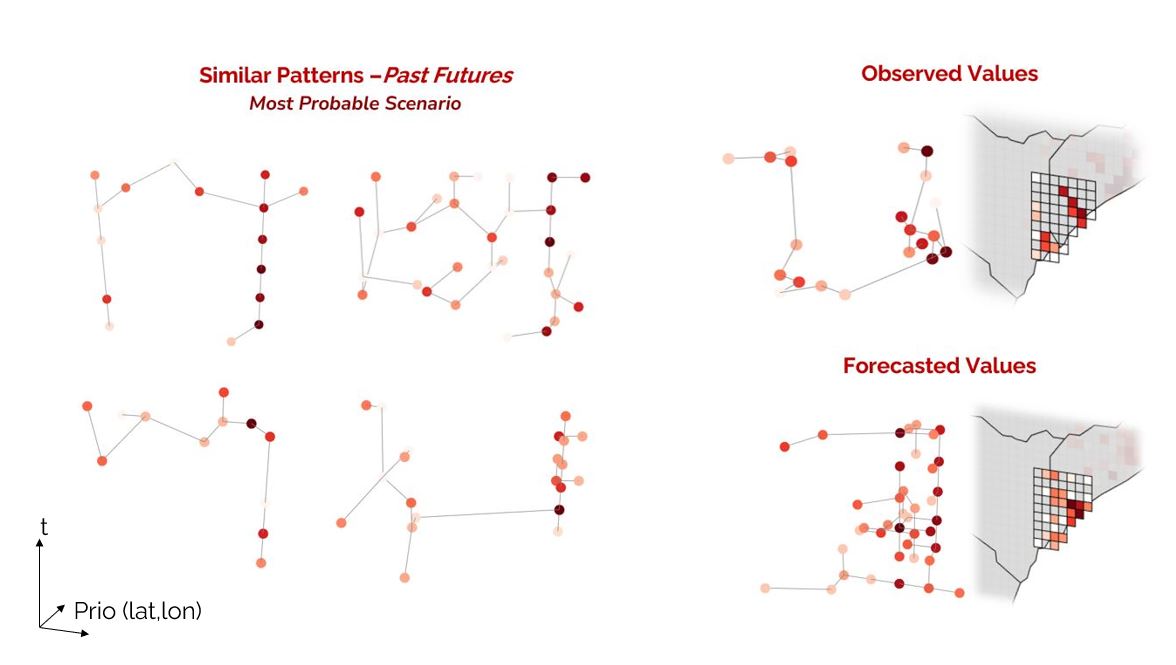}
\caption{Outcome of the ShapeFinder model for the Kenya-Somalia example, discussed in Figure \ref{expla_test}. Four examples of past futures, and the corresponding Forecasted and Observed values. The left side of the figure displays four ``following values'' examples (also referred to as past futures) of similar historical sequences. The right side of the figure represents the forecasted and actual observed values. The 3D sequence uses longitude on the x-axis, latitude on the y-axis, and time on the z-axis. Each point represents the fatality value for a Prio-Grid cell in a specific month. Darker red points indicate higher fatality values, while gray links between points assist in visualizing the 3D structure but carry no interpretative meaning. Accompanying each 3D sequence is a map with accumulated fatalities over the period per Prio-Grid cell. The black grid represents the event’s location, and the colored squares indicate the total fatalities accumulated during the entire period within the Prio-Grid cells.}
\label{kenya_outcome}
\end{figure}

Figure \ref{kenya_outcome} illustrates the Kenya-Somalia example, discussed in Figure \ref{expla_test}. The left side of the figure displays four ``following values'' examples (also referred to as past futures) of similar historical sequences. These examples are extracted from the most probable scenario, which includes ten sequences in total. The right side of the figure compares the forecasted and actual observed values. The forecasted values are constructed with the following values of similar historical sequences shown on the left. This pattern's occurrences share a common dynamic: most fatalities are concentrated in a couple of Prio-Grid cells, located on the right side of the sequence. The sequences also have a secondary zone of fatalities, more minor. In some cases, this secondary zone is concentrated within one or two cells on the left part, while in others, it appears more diffusely distributed around the primary zone of concentration.

The resulting forecast mirrors the observed values to a significant extent, correctly predicting the concentration of fatalities in two cells along the coast near the border. However, while the observed values reveal a cluster of fatalities in the south, this area is slightly underpredicted in the prediction. Instead, the model forecasts a more diffuse spread of fatalities around the two primary cells, failing to capture the precise secondary cluster. More examples of forecasted projections of fatalities from the SF model and the Views model against the actual observed values are included in Figure \ref{app:ex_patterns}.

\subsection*{Meaningful patterns ?}

A shape-based forecasting model operates on the assumption that similar occurrences of a pattern lead to similar outcomes. To evaluate this assumption, we examine whether the similarity of past futures between two sequences correlates with their distance. For this, we analyze all sequences in the study and compare them against historically recorded sequences that share comparable dimensions and active cell counts. We measure both the sequences' similarity metric EMD ($EMD_{pattern}$) but also the differences in their ratio of increase ($\Delta R_{inc}$) and EMD for their past futures ($EMD_{pf}$).

\begin{figure}[!h]
\centering
\includegraphics[width=\textwidth]{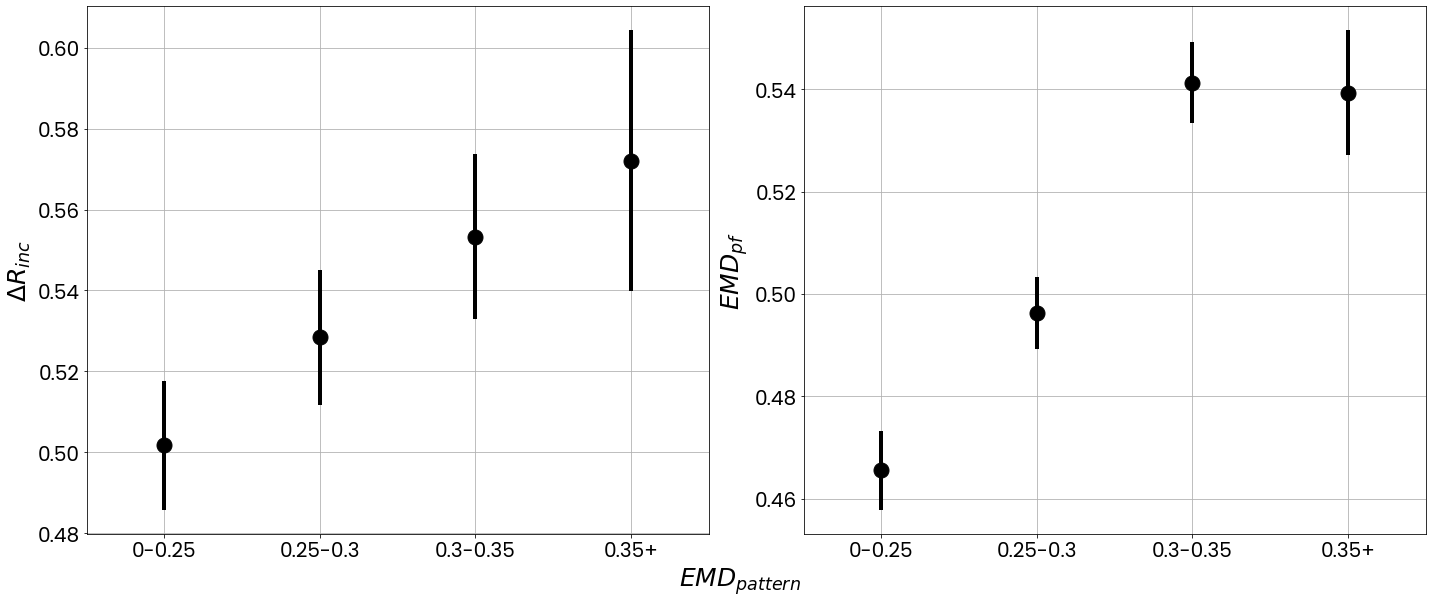}
\caption{The mean and 95\% confidence interval for sequences' similarity metric $EMD_{pattern}$ (x-axis) and the differences in their ratio of increase $\Delta R_{inc}$ on the left part and EMD for their past futures $EMD_{pf}$ on the right (y-axis). The confidence interval is computed as the standard error divided by the square root of the number of observations.}
\label{match_fut_val}
\end{figure}

Figure \ref{match_fut_val} presents the results, where values are binned for clarity. The mean and 95\% confidence interval for each bin are displayed for $R_{inc}$ (left) and $EMD_{pf}$ (right). Both $R_{inc}$ and $EMD_{pf}$ of the smaller bin are significantly smaller (except the 0.25-0.3 bin for $\Delta R_{inc}$). Generally, when $EMD_{pattern}$ increases, $R_{inc}$ and $EMD_{pf}$ increases too. It confirms that more similar patterns tend to produce more similar outcomes (past futures), validating the model's core assumption.

\subsection*{Prediction at the Prio-Grid level}

First, the Prio-Grid cell is the unit of analysis. In this section, we compare the model results at the micro-level, providing insight into the prediction performance at the smallest geographical grain of analysis. 

\begin{figure}[!h]
\centering
\includegraphics[width=0.9\textwidth]{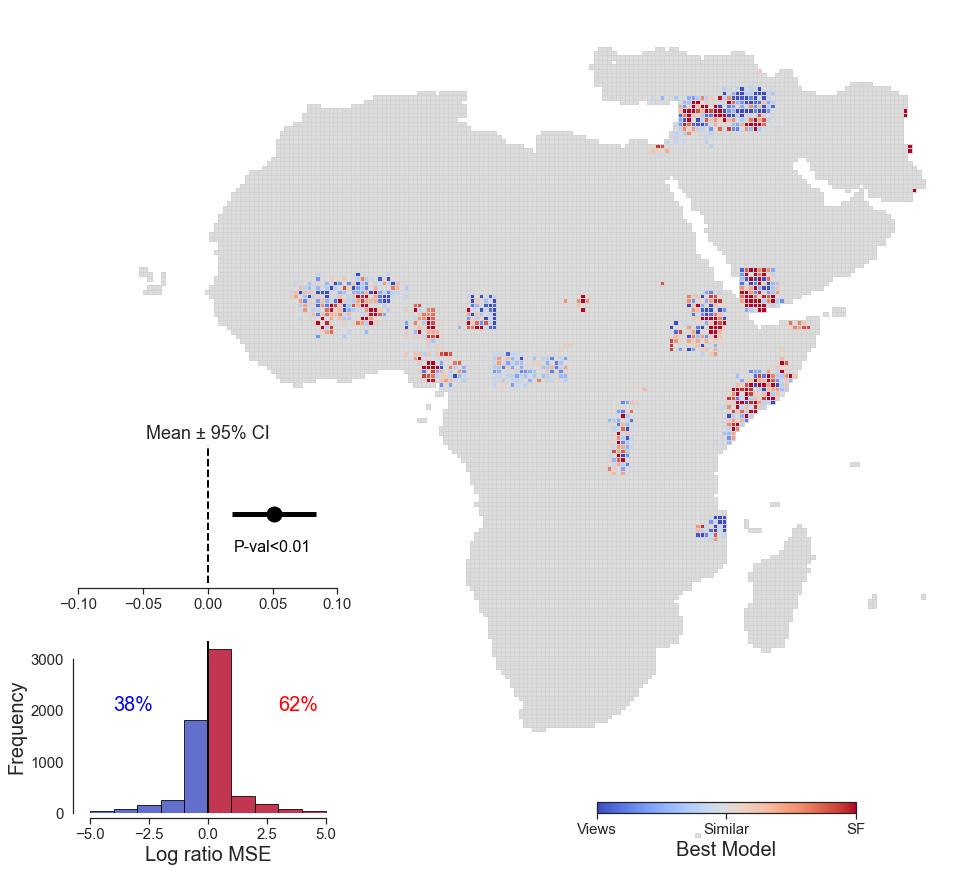}  
\caption{Log ratio of Mean Squared Error (MSE) between the Views model and the ShapeFinder (SF) model at the Prio-Grid level and Mean and 95\% of the Confidence level, calculated with the standard error divided by the square root of the number of observations. The units in the two subfigures (bottom left of the figure) are the metrics calculated for the 6-month prediction period and Prio-Grid. Along with this, the map (center of the figure) displays cumulative MSE per cell over four analysis periods. Red indicates better performance by the SF model, blue by the Views model, and gray denotes similar predictions.}
\label{mse_grid}
\end{figure}

Figure \ref{mse_grid} shows the log ratio of Mean Squared Error (MSE) between the Views model and the ShapeFinder (SF) model at the Prio-Grid level. The units in the two subfigures (bottom left of the figure) are the metrics calculated for the 6-month prediction period and Prio-Grid. Across the four periods analyzed, approximately 1,500 Prio-Grid cells per period (totaling just over 6,000 cells) are included. A positive ratio (in red) indicates that the ShapeFinder model outperforms the Views model. Overall, the SF model demonstrates significantly better performance, with an average log ratio of 0.05. The median log ratio is positive, and in total, the SF model shows improvements in over 62\% of cases. However, many of these improvements are minor, as indicated by the substantial number of points included in the histogram bin from 0 to 1. Along with this, a map displays cumulative MSE per cell over four analysis periods. Red indicates better performance by the SF model, blue by the Views model, and gray denotes similar predictions. The Views model performs better in northern Iraq, the Central African Republic, and the Chad-Nigeria border. Mixed results appear in Burkina Faso, the DRC-Rwanda border, and Somalia. Finally, ShapeFinder outperforms in Yemen, Nigeria, central Ethiopia, and Syria. \ref{app:add_res} presents the detailed error metrics for each forecast horizon and test period, and \ref{app:prio} provides examples of the model’s predictions.

\subsection*{Prediction at the Zone level}

In complement, this section focuses on the macro-level results by analyzing the model’s performance at the zone level, as discussed in Figure \ref{ex_pattern}. This approach is more relevant for early warning applications, as it considers entire zones rather than isolated grid cells. By capturing broader conflict dynamics, this level of analysis provides a more meaningful foundation for decision-making and risk assessment.

\begin{figure}[!h]
\centering
\includegraphics[width=0.8\textwidth]{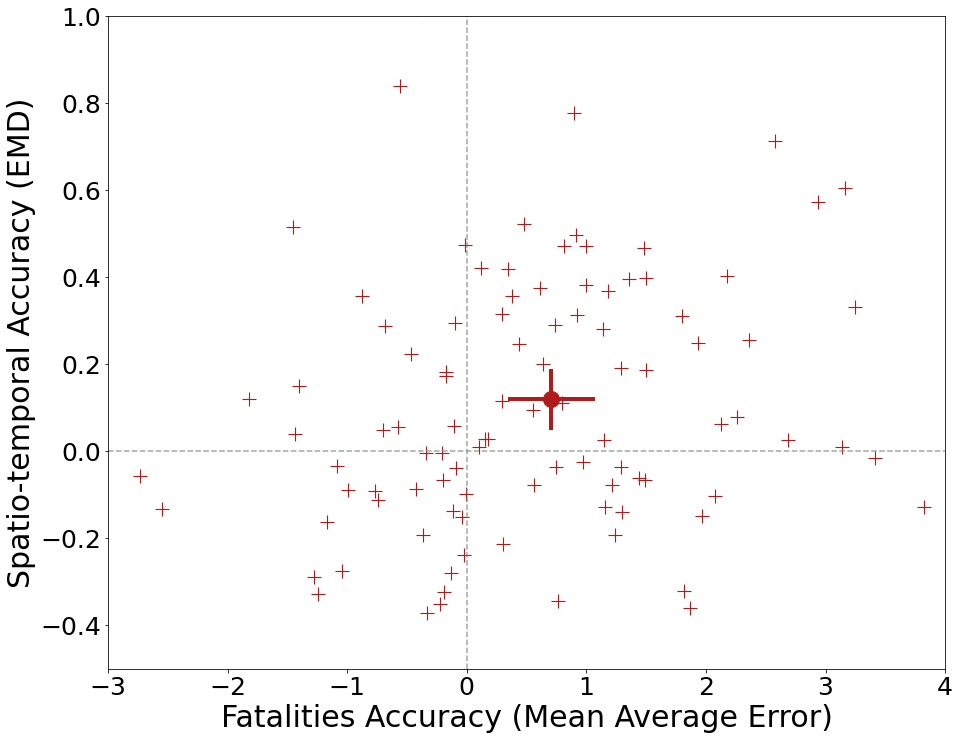}  \caption{Scatter plot of the log ratio of Raw Error (in x) and EMD (in y) at the zone level, as introduced in Figure \ref{ex_pattern}, of Views model over ShapeFinder model. If the ratio is positive, the ShapeFinder model shows better performance. The red point is the mean value with the 95\% Confidence Interval, calculated with the standard error divided by the square root of the number of observations.}
\label{main_metric}
\end{figure}

Figure \ref{main_metric} presents the log ratio of Mean Average Error (MAE) (x-axis) and EMD (y-axis) at the pattern level, comparing the Views model to the ShapeFinder (SF) model. A positive ratio indicates that the SF model outperforms the Views model. The results demonstrate that the SF model outperforms in both MAE (average log ratio = 0.708) and EMD (average log ratio = 0.120), achieving the most accurate predictions for total fatalities and dynamic patterns. In total, 111 patterns are included across the four periods. Appendix \ref{app:pattern} shows examples of projections and forecast maps to further illustrate these findings.

\subsection*{Performance analysis}

To assess where and why the SF model outperforms, we examine three characteristics of the 111 identified active zones across the four studied periods. First, we analyze how an input sequence evolves in terms of the intensity of fatalities using the ratio of increase $R_{inc}$. Second, we evaluate two performance measures: the prediction performance of Views (using the Mean Absolute Percentage Error, MAPE) and the relative difference in performance between Views and the SF model (log ratio of error). These metrics allow us to identify regions where Views performs poorly and to determine whether the SF model minimizes these errors, showing clearly areas where the SF model contributes improvements.

\begin{figure}[!h]
\centering
\includegraphics[width=0.7\textwidth]{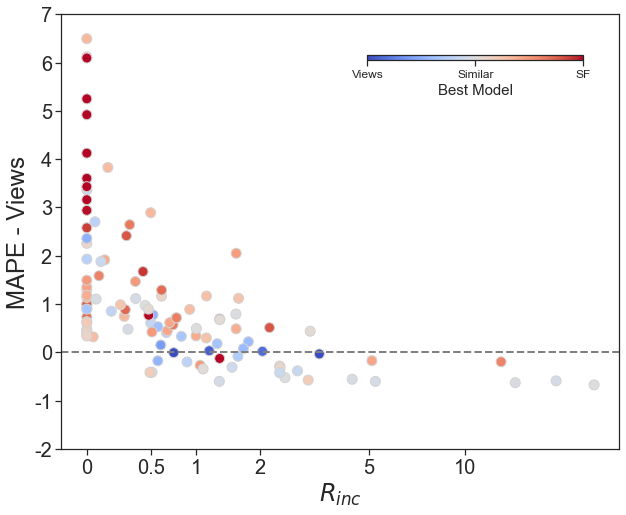}  \caption{Scatter plot of the ratio of increase $R_{inc}$ (in x) and Mean Average Percentage Error (MAPE) of Views (in y) at the active zone level, with a log-modulus transformation. The color of the point represents the difference in performance between the two models. The darker red, the SF performs better, and the darker blue, the Views model.}
\label{expla_why}
\end{figure}

Figure \ref{expla_why} illustrates the relationship between the input sequence’s increase ratio $R_{inc}$ (x-axis), the Mean Average Percentage Error (MAPE) of the Views model (y-axis) with a log-modulus transformation, and the error ratio of both models (indicated by color). In the figure, darker red areas show where the SF model performs better, while darker blue areas indicate better performance by the Views model. As expected, inputs where the Views model has strong MAPE performances (around the 0-line on the y-axis) also show a negative error ratio (in blue), indicating better performance for Views. Beyond this, we can see three main groups of inputs. First, there is a group of dark red points with $R_{inc}$ value from 0 to 0.5, where the Views model overpredicts significantly (with MAPE values above 2). The second group includes light red points with $R_{inc}$ value up to 2, where Views slightly overpredicts. Finally, there is a group of very pale red and blue points with $R_{inc}$ value above 3, where both models underpredict (negative MAPE values). These are the hardest patterns to forecast, as they involve sudden, large increases in fatalities, and both models underpredict similarly here. Overall, it appears that the SF model’s improvement comes mainly in cases where the Views model overpredicts.

An important question is then whether the SF model achieves its strong MSE and Absolute Error scores simply by making conservative predictions. Metrics like these can favor models that produce conservative prediction values, especially in conflict forecasting at a fine scale (monthly, grid-level), where most values are zero. However, Figures \ref{app:glob_over} and \ref{app:max_over} show the MAPE and Maximum Error, respectively. These reveal that the SF model strikes a balance between under- and over-predictions. Interestingly, its maximum forecast values are even higher in mean and median than those of the Views model. This suggests that the SF model provides more accurate and higher-scoring predictions without relying on overly conservative estimates.

\begin{figure}[!h]
\centering
\includegraphics[width=\textwidth]{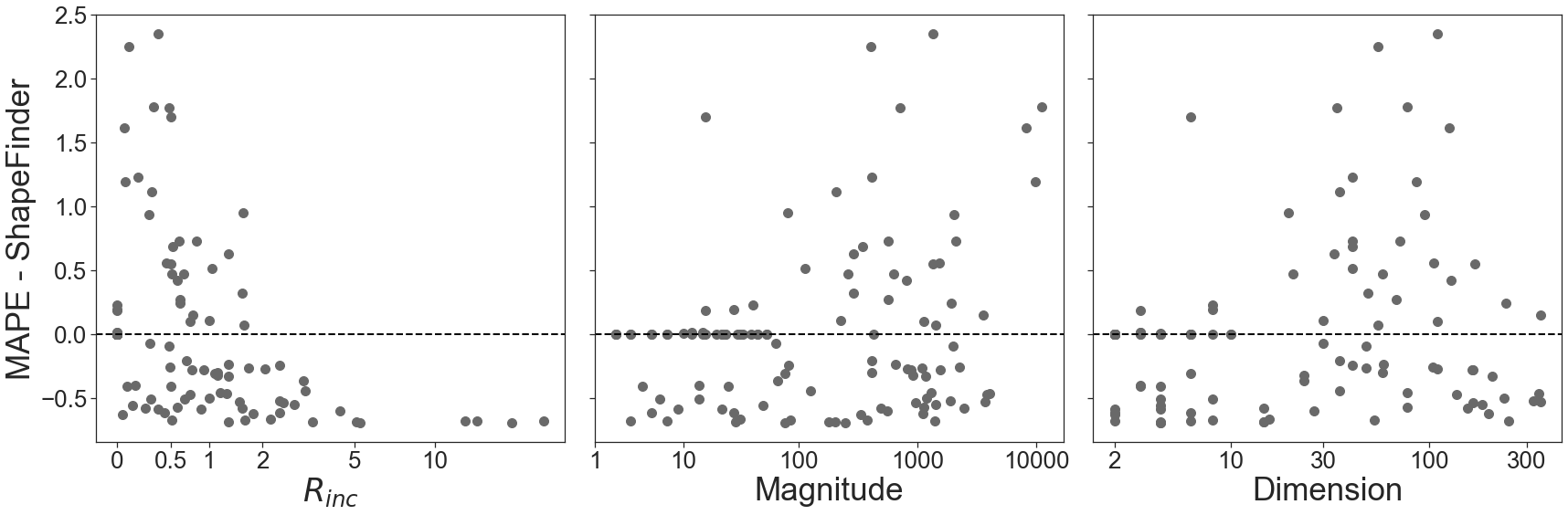}  \caption{Scatter plots showing the relationship between input sequence characteristics and prediction error at the zone level using the ShapeFinder model. The y-axis in all three plots represents the Mean Absolute Percentage Error (MAPE), with a log-modulus transformation, where positive values indicate overpredictions and negative values indicate underpredictions. The x-axes represent: (left) the ratio of increase 
$R_{inc}$, (center) the total number of fatalities in the input pattern (referred to as Magnitude), and (right) the number of active cells in the pattern (referred to as Dimension). The dotted horizontal line at 0 on the y-axis indicates a perfect prediction.}
\label{expla_why_sf}
\end{figure}

More detailed plots of the SF performances are displayed in Figure \ref{expla_why_sf}. It shows the relationship between pattern characteristics and prediction error at the pattern level using the ShapeFinder model. The y-axis in all three plots represents the Mean Absolute Percentage Error (MAPE), with a log-modulus transformation, where positive values indicate overpredictions and negative values indicate underpredictions. The x-axes represent: (left) the ratio of increase 
$R_{inc}$, (center) the total number of fatalities in the input pattern (referred to as Magnitude), and (right) the number of active cells in the pattern (referred to as Dimension). The dotted horizontal line at 0 on the y-axis indicates a perfect prediction. Similarly to the Views, all of the overpredictions are situated when $R_{inc}$ is inferior to 2. However, they also correspond to high magnitude input patterns and medium dimension size patterns (from 30 to 100 active cells). The underpredictions are more uniformly distributed and less important in values than overpredictions.

\section{Discussion}

In this paper, we introduce a novel method for identifying conflict dynamics at a fine-grained geographic level. It shows strong forecasting performances, overperforming the Views model, one of the strongest benchmarks, at both micro and macro levels. A deeper analysis of our model’s performance shows that it might offer a more precise forecasted dynamic. It seems that the Views model captures the general area where fatalities might occur but lacks precision within that area, often leading to overpredictions. The EMD metric supports this, indicating that our model better pinpoints specific locations. This difference is also clear in Figure \ref{app:ex_patterns}, where aggregated forecasts show that our model provides a more refined pattern, while Views predicts broad, vague clouds of activity. But, as noted in the Benchmark section, the Views model is optimized for a 36-month forecast horizon. If it were optimized for a shorter 6-month horizon, its results might differ significantly. 

However, this method is based on several assumptions. The primary assumption is that patterns of different magnitudes in geography and intensity are comparable. While studies have shown that this approach works at larger scales, such as country-month analyses \citep{schincariol2025accounting}, its effectiveness at smaller scales, as explored in this study, requires further investigation\footnote{The results were also compared with the original method, showing better forecasting performances. The results are presented in \ref{app:comp_SF}}. Future work could adopt a more theory-driven approach, using the model to examine patterns associated with specific variables such as conflict type or commonly used covariates in conflict modeling. Moreover, while this study is limited to Africa and the Middle East, future research could explore cross-regional pattern similarity and extend the model to a broader geographic scope. 

In addition to this future innovation, another possible direction is to move away from the Prio-Grid system. UCDP provides information on geographic accuracy, which would allow the model to retain only events with high spatial precision. However, removing imprecise events introduces a new problem: if the excluded events involve large numbers of fatalities, their absence could substantially change the overall shape and produce an entirely different pattern. An alternative approach would be to keep the best estimate of the precise location and replace the cubic density measure with a cylindrical density measure, using only latitude and longitude as the geographic dimensions. But similarly, the geographic imprecision of a heavy fatalities event might alter the pattern shape.

The prediction procedure itself can also be adapted in future work. The conflict forecasting field is moving away from point estimates \citep{randahl2024bin}, as illustrated by the most recent Views competition \citep{hegre20242023}, where participants were required to submit full probability distributions rather than single-value forecasts. The ShapeFinder can be adapted in this direction by shifting from scenario generation to estimating a predictive distribution for each cell-month, based on the future values observed in similar historical cases. Further research is needed to determine both the appropriate statistical distribution (the country-month forecasts in the competition used a Poisson distribution) and the optimal balance between the number of similar cases required and the distance thresholds used to classify two 3D sequences as similar ($thr1$ and $thr2$ in Figure \ref{recap}). Fitting a full distribution will likely require a larger set of similar historical sequences, which may affect the choice of these thresholds.

Besides significant strengths, the model faces some limitations. First, even if the model is highly flexible, it assumes that the diffusion of conflict to neighboring areas is equally probable. In reality, this is not the case, as factors such as logistics (e.g., roads connecting cells) and human communication networks, which might link distant neighbors more closely than geographically closer ones due to differences in religion, culture, etc., play a significant role. Addressing these factors would require a major overhaul of the model, including the incorporation of road networks and population data for each cell. This would add considerable complexity to the model and necessitate a large amount of data. 

Then, a key advantage of the original ShapeFinder model that is reduced in this adaptation is its strong interpretability. The two-dimensional country-level patterns make it possible to draw conclusions about how a specific shape might lead to outcome y or z, since the structure of the shape can be directly examined. In the model presented in this paper, which relies on three dimensions, and even four if we include fatalities as an additional dimension, this interpretability is strongly limited. Humans struggle to intuitively understand three-dimensional shapes, especially when they are projected onto a flat page, making it difficult to see what the shape represents or to judge its similarity to others. The only aspect of interpretability that remains clear in this version is traceability: the ability to identify where and when similar patterns occurred in the historical data. This still enables meaningful contextual analysis, even if the shapes themselves are harder to interpret.

Another limitation of our model is that, because the model is fully autoregressive, it cannot forecast conflict in zones where no conflict has occurred in the recent past. However, unlike the original country-level ShapeFinder, where the inability to predict onsets was an explicit limitation, this constraint is only partially applicable here. Cells that are currently non-active but fall within an active zone can still be forecasted as non-zero, even if they recorded only zeros in the previous months. In this sense, the current version of ShapeFinder resembles a hurdle model, similar to the approach in \cite{fritz2024predicting}. The difference is that, instead of using a separate model to identify which zones require forecasting, we include all zones that recorded fatalities in the previous 12 months. As a result, the inability to forecast onsets applies only to cells outside these active zones. 

The results included in \ref{imp_acs} show that this zone detection method allows for capturing most of the future fatalities, confirming the purely autoregressive compatibility for active conflict, meaning that active conflict seems to remain in the same active zones. However, we also see that the onsets of fatalities corresponding to around 10\% of the fatalities are impossible to capture. Predicting these spikes is challenging for all models, even those with additional data. This is clear in Figure \ref{expla_why}, where the Views model underpredicts values when the increase ratio is above 3. In those zones, predicting zero remains the best strategy for optimizing the error metrics (see Figure \ref{zero_test}). 

This raises a question of the useful metrics to assess our models. A general model optimized on the error is unlikely to predict an onset of conflict, as the reward is too low compared to the error risk of forecasting such a rare event. However, a key expectation of forecasting models from decision-makers is to detect those early warning signals to ideally avoid conflict onsets. The solution could be to develop hurdle models that are assessed on both error in active conflict zones and a highly rewarding onset metric for non-active zones. 

\subsection*{Conclusion}
Our model demonstrates strong forecasting performance with the added benefit of transparency. Its main asset is the traceability of the source of each forecast. Contrary to highly complex machine learning models, our approach allows us to trace the origins of each forecast, showing when similar patterns occurred in the past and the likelihood of different future scenarios. This makes the model transparent, an important advantage for convincing decision-makers of its reliability. Moreover, it can run quickly (about one hour per period) compared to the much longer processing times of deep learning methods, and only needs fatality input data to run. 

\clearpage
\section*{Acknowledgements}
This project has received funding from the European Research Council (ERC) under the European Union's Horizon 2020 research and innovation programme (grant agreement No 101002240)

\section*{CRediT authorship contribution statement}
Thomas Schincariol: Conceptualization; Data curation; Formal analysis; Methodology; Software; Validation; Visualization; Writing - original draft; and Writing - review \& editing

\clearpage
\section*{Declaration of competing interest}
The authors declare that they have no competing interests.

\section*{Declaration of Generative AI in the writing process}

I used ChatGPT to assist with grammar and spelling corrections, translations, and synonym suggestions during the preparation of this work. I reviewed and edited all content afterward and take full responsibility for the final manuscript.

\clearpage
\bibliographystyle{apsr}
\bibliography{biblio}

@unpublished{acled_cast_2025,
	title = {ACLED Conflict Alert System (CAST)},
	author = {ACLED},
	year = {2025},
    note={Armed Conflict Location and Event Data (ACLED). Available at: \href{https://acleddata.com/platform/cast-conflict-alert-system}{https://acleddata.com/platform/cast-conflict-alert-system} (accessed 9 December 2025).},
}

@article{randahl2024bin,
  title={Bin-Conditional Conformal Prediction of Fatalities from Armed Conflict},
  author={Randahl, David and Williams, Jonathan P and Hegre, H{\aa}vard},
  journal={arXiv preprint arXiv:2410.14507},
  year={2024}
}

@article{weidmann2015accuracy,
  title={On the accuracy of media-based conflict event data},
  author={Weidmann, Nils B},
  journal={Journal of Conflict Resolution},
  volume={59},
  number={6},
  pages={1129--1149},
  year={2015},
  publisher={SAGE Publications Sage CA: Los Angeles, CA}
}

@misc{fritz2024predicting,
  title={Predicting uncertainty in stages: Using a semiparametric hierarchical hurdle model for predicting distributions of conflict fatalities},
  author={Fritz, Cornelius and Dworschak, Christoph and Mehrl, Marius},
  year={2024},
  publisher={June}
}

@article{mueller2024introducing,
  title={Introducing a global dataset on conflict forecasts and news topics},
  author={Mueller, Hannes and Rauh, Christopher and Seimon, Ben},
  journal={Data \& Policy},
  volume={6},
  pages={e17},
  year={2024},
  publisher={Cambridge University Press}
}

@article{shwartz2022tabular,
  title={Tabular data: Deep learning is not all you need},
  author={Shwartz-Ziv, Ravid and Armon, Amitai},
  journal={Information Fusion},
  volume={81},
  pages={84--90},
  year={2022},
  publisher={Elsevier}
}

@article{fayaz2022deep,
  title={Is deep learning on tabular data enough? An assessment},
  author={Fayaz, Sheikh Amir and Zaman, Majid and Kaul, Sameer and Butt, Muheet Ahmed},
  journal={International Journal of Advanced Computer Science and Applications},
  volume={13},
  number={4},
  pages={466--473},
  year={2022},
  publisher={Science and Information (SAI) Organization Limited}
}

@article{hegre20242023,
  title={The 2023/24 VIEWS Prediction challenge: Predicting the number of fatalities in armed conflict, with uncertainty},
  author={Hegre, H{\aa}vard and others},
  journal={Journal of Peace Research},
  pages={00223433241300862},
  year={2024},
  publisher={SAGE Publications Sage UK: London, England}
}

@article{zammit2012point,
  title={Point process modelling of the Afghan War Diary},
  author={Zammit-Mangion, Andrew and Dewar, Michael and Kadirkamanathan, Visakan and Sanguinetti, Guido},
  journal={Proceedings of the National Academy of Sciences},
  volume={109},
  number={31},
  pages={12414--12419},
  year={2012},
  publisher={National Academy of Sciences}
}

@article{cook2023stadl,
  title={Stadl up! the spatiotemporal autoregressive distributed lag model for tscs data analysis},
  author={Cook, Scott J and Hays, Jude C and Franzese Jr, Robert J},
  journal={American Political Science Review},
  volume={117},
  number={1},
  pages={59--79},
  year={2023},
  publisher={Cambridge University Press}
}

@article{davis2024fractional,
  title={A fractional Hawkes process model for earthquake aftershock sequences},
  author={Davis, Louis and Baeumer, Boris and Wang, Ting},
  journal={Journal of the Royal Statistical Society Series C: Applied Statistics},
  volume={73},
  number={5},
  pages={1185--1202},
  year={2024},
  publisher={Oxford University Press UK}
}

@article{reinhart2018self,
  title={Self-exciting point processes with spatial covariates: modelling the dynamics of crime},
  author={Reinhart, Alex and Greenhouse, Joel},
  journal={Journal of the Royal Statistical Society Series C: Applied Statistics},
  volume={67},
  number={5},
  pages={1305--1329},
  year={2018},
  publisher={Oxford University Press}
}

@article{yu2017spatio,
  title={Spatio-temporal graph convolutional networks: A deep learning framework for traffic forecasting},
  author={Yu, Bing and Yin, Haoteng and Zhu, Zhanxing},
  journal={arXiv preprint arXiv:1709.04875},
  year={2017}
}

@article{corso2024graph,
  title={Graph neural networks},
  author={Corso, Gabriele and Stark, Hannes and Jegelka, Stefanie and Jaakkola, Tommi and Barzilay, Regina},
  journal={Nature Reviews Methods Primers},
  volume={4},
  number={1},
  pages={17},
  year={2024},
  publisher={Nature Publishing Group UK London}
}

@article{reinhart2018review,
  title={A review of self-exciting spatio-temporal point processes and their applications},
  author={Reinhart, Alex},
  journal={Statistical Science},
  volume={33},
  number={3},
  pages={299--318},
  year={2018},
  publisher={JSTOR}
}

@article{hawkes1971spectra,
  title={Spectra of some self-exciting and mutually exciting point processes},
  author={Hawkes, Alan G},
  journal={Biometrika},
  volume={58},
  number={1},
  pages={83--90},
  year={1971},
  publisher={Oxford University Press}
}

@article{racek2025capturing,
  title={Capturing the spatio-temporal diffusion effects of armed conflict: A nonparametric smoothing approach},
  author={Racek, Daniel and Thurner, Paul W and Kauermann, G{\"o}ran},
  journal={Journal of the Royal Statistical Society Series A: Statistics in Society},
  pages={qnaf120},
  year={2025},
  publisher={Oxford University Press UK}
}

@article{von2025next,
  title={Next-Generation Conflict Forecasting: Unleashing Predictive Patterns through Spatiotemporal Learning},
  author={von der Maase, Simon P},
  journal={arXiv preprint arXiv:2506.14817},
  year={2025}
}

@article{schincariol2025accounting,
  title={Accounting for variability in conflict dynamics: A pattern-based predictive model},
  author={Schincariol, Thomas and Frank, Hannah and Chadefaux, Thomas},
  journal={Journal of Peace Research},
  pages={00223433251330790},
  year={2025},
  publisher={SAGE Publications Sage UK: London, England}
}

@article{cook2022race,
  title={Race to the bottom: Spatial aggregation and event data},
  author={Cook, Scott J and Weidmann, Nils B},
  journal={International Interactions},
  volume={48},
  number={3},
  pages={471--491},
  year={2022},
  publisher={Taylor \& Francis}
}

@misc{walterskirchen2024taking,
  title={Taking time seriously: Predicting conflict fatalities using temporal fusion transformers},
  author={Walterskirchen, Julian and H{\"a}ffner, Sonja and Oswald, Christian and Binetti, Marco Nicola},
  year={2024},
  publisher={SocArXiv}
}

@article{radford2022high,
  title={High resolution conflict forecasting with spatial convolutions and long short-term memory},
  author={Radford, Benjamin J},
  journal={International Interactions},
  volume={48},
  number={4},
  pages={739--758},
  year={2022},
  publisher={Taylor \& Francis}
}

@article{kathman2010civil,
  title={Civil war contagion and neighboring interventions},
  author={Kathman, Jacob D},
  journal={International Studies Quarterly},
  volume={54},
  number={4},
  pages={989--1012},
  year={2010},
  publisher={Blackwell Publishing Ltd Oxford, UK}
}

@article{black2013have,
  title={When have violent civil conflicts spread? Introducing a dataset of substate conflict contagion},
  author={Black, Nathan},
  journal={Journal of Peace Research},
  volume={50},
  number={6},
  pages={751--759},
  year={2013},
  publisher={SAGE Publications Sage UK: London, England}
}

@article{cederman2017predicting,
  title={Predicting armed conflict: Time to adjust our expectations?},
  author={Cederman, Lars-Erik and Weidmann, Nils B},
  journal={Science},
  volume={355},
  number={6324},
  pages={474--476},
  year={2017},
  publisher={American Association for the Advancement of Science}
}

@article{kelling2020analysis,
  title={Analysis of Conflict Diffusion Over Continuous Space},
  author={Kelling, Claire and Lin, YiJyun},
  journal={Computational Conflict Research},
  pages={201--223},
  year={2020},
  publisher={Springer International Publishing}
}

@article{schutte2011diffusion,
  title={Diffusion patterns of violence in civil wars},
  author={Schutte, Sebastian and Weidmann, Nils B},
  journal={Political Geography},
  volume={30},
  number={3},
  pages={143--152},
  year={2011},
  publisher={Elsevier}
}

@inproceedings{li2024urbangpt,
  title={Urbangpt: Spatio-temporal large language models},
  author={Li, Zhonghang and Xia, Lianghao and Tang, Jiabin and Xu, Yong and Shi, Lei and Xia, Long and Yin, Dawei and Huang, Chao},
  booktitle={Proceedings of the 30th ACM SIGKDD Conference on Knowledge Discovery and Data Mining},
  pages={5351--5362},
  year={2024}
}

@article{aas2011all,
  title={All conflict is local: Modeling sub-national variation in civil conflict risk},
  author={Aas Rustad, Siri Camilla and Buhaug, Halvard and Falch, {\AA}shild and Gates, Scott},
  journal={Conflict Management and Peace Science},
  volume={28},
  number={1},
  pages={15--40},
  year={2011},
  publisher={SAGE Publications Sage UK: London, England}
}

@inproceedings{zhang2024large,
  title={Large language models for spatial trajectory patterns mining},
  author={Zhang, Zheng and Amiri, Hossein and Liu, Zhenke and Zhao, Liang and Z{\"u}fle, Andreas},
  booktitle={Proceedings of the 1st ACM SIGSPATIAL International Workshop on Geospatial Anomaly Detection},
  pages={52--55},
  year={2024}
}

@inproceedings{andoni2008earth,
  title={Earth mover distance over high-dimensional spaces.},
  author={Andoni, Alexandr and Indyk, Piotr and Krauthgamer, Robert},
  booktitle={SODA},
  volume={8},
  pages={343--352},
  year={2008}
}

@article{tollefsen2012prio,
  title={PRIO-GRID: A unified spatial data structure},
  author={Tollefsen, Andreas For{\o} and Strand, H{\aa}vard and Buhaug, Halvard},
  journal={Journal of Peace Research},
  volume={49},
  number={2},
  pages={363--374},
  year={2012},
  publisher={Sage Publications Sage UK: London, England}
}

@article{lindholm2022predicting,
  title={Predicting political violence using a state-space model},
  author={Lindholm, Andreas and Hendriks, Johannes and Wills, Adrian and Sch{\"o}n, Thomas B},
  journal={International Interactions},
  volume={48},
  number={4},
  pages={759--777},
  year={2022},
  publisher={Taylor \& Francis}
}

@article{vesco2022united,
  title={United they stand: Findings from an escalation prediction competition},
  author={Vesco, Paola and Hegre, H{\aa}vard and Colaresi, Michael and Jansen, Remco Bastiaan and Lo, Adeline and Reisch, Gregor and Weidmann, Nils B},
  journal={International Interactions},
  volume={48},
  number={4},
  pages={860--896},
  year={2022},
  publisher={Taylor \& Francis}
}

@article{dorussen2016networked,
  title={Networked international politics: Complex interdependence and the diffusion of conflict and peace},
  author={Dorussen, Han and Gartzke, Erik A and Westerwinter, Oliver},
  journal={Journal of peace research},
  volume={53},
  number={3},
  pages={283--291},
  year={2016},
  publisher={SAGE Publications Sage UK: London, England}
}

@techreport{racek2024integrating,
  title={Integrating Spatio-temporal Diffusion into Statistical Forecasting Models of Armed Conflict via Non-parametric Smoothing},
  author={Racek, Daniel and Thurner, Paul and Kauermann, Goeran},
  year={2024},
  institution={Center for Open Science}
}

@article{bazzi2022promise,
  title={The promise and pitfalls of conflict prediction: evidence from Colombia and Indonesia},
  author={Bazzi, Samuel and Blair, Robert A and Blattman, Christopher and Dube, Oeindrila and Gudgeon, Matthew and Peck, Richard},
  journal={Review of Economics and Statistics},
  volume={104},
  number={4},
  pages={764--779},
  year={2022},
  publisher={MIT Press One Rogers Street, Cambridge, MA 02142-1209, USA journals-info~…}
}

@article{fritz2022role,
  title={The role of governmental weapons procurements in forecasting monthly fatalities in intrastate conflicts: A semiparametric hierarchical hurdle model},
  author={Fritz, Cornelius and Mehrl, Marius and Thurner, Paul W and Kauermann, G{\"o}ran},
  journal={International Interactions},
  volume={48},
  number={4},
  pages={778--799},
  year={2022},
  publisher={Taylor \& Francis}
}

@article{hegre2019views,
  title={ViEWS: A political violence early-warning system},
  author={Hegre, H{\aa}vard and Allansson, Marie and Basedau, Matthias and Colaresi, Michael and Croicu, Mihai and Fjelde, Hanne and Hoyles, Frederick and Hultman, Lisa and H{\"o}gbladh, Stina and Jansen, Remco and others},
  journal={Journal of peace research},
  volume={56},
  number={2},
  pages={155--174},
  year={2019},
  publisher={SAGE Publications Sage UK: London, England}
}

@article{racek2024conflict,
  title={Conflict forecasting using remote sensing data: An application to the Syrian civil war},
  author={Racek, Daniel and Thurner, Paul W and Davidson, Brittany I and Zhu, Xiao Xiang and Kauermann, G{\"o}ran},
  journal={International Journal of Forecasting},
  volume={40},
  number={1},
  pages={373--391},
  year={2024},
  publisher={Elsevier}
}

@article{iqbal2008bad,
  title={Bad neighbors: Failed states and their consequences},
  author={Iqbal, Zaryab and Starr, Harvey},
  journal={Conflict Management and Peace Science},
  volume={25},
  number={4},
  pages={315--331},
  year={2008},
  publisher={Taylor \& Francis}
}

@article{metternich2017firewall,
  title={Firewall? Or wall on fire? A unified framework of conflict contagion and the role of ethnic exclusion},
  author={Metternich, Nils W and Minhas, Shahryar and Ward, Michael D},
  journal={Journal of Conflict Resolution},
  volume={61},
  number={6},
  pages={1151--1173},
  year={2017},
  publisher={SAGE Publications Sage CA: Los Angeles, CA}
}

@article{buhaug2008contagion,
  title={Contagion or confusion? Why conflicts cluster in space},
  author={Buhaug, Halvard and Gleditsch, Kristian Skrede},
  journal={International studies quarterly},
  volume={52},
  number={2},
  pages={215--233},
  year={2008},
  publisher={Blackwell Publishing Oxford, UK}
}

@article{davies2022organized,
  title={Organized violence 1989--2021 and drone warfare},
  author={Davies, Shawn and Pettersson, Therese and {\"O}berg, Magnus},
  journal={Journal of Peace Research},
  volume={59},
  number={4},
  pages={593--610},
  year={2022},
  publisher={SAGE Publications Sage UK: London, England}
}

@article{sundberg2013introducing,
  title={Introducing the UCDP georeferenced event dataset},
  author={Sundberg, Ralph and Melander, Erik},
  journal={Journal of Peace Research},
  volume={50},
  number={4},
  pages={523--532},
  year={2013},
  publisher={Sage Publications Sage UK: London, England}
}

@misc{hegre_forecasting_2022,
	title = {Forecasting fatalities in armed conflict: {Forecasts} for {April} 2022-{March} 2025},
	author = {Hegre, Håvard and Lindqvist-McGowan, Angelica and Dale, James and Croicu, Mihai and Randahl, David and Vesco, Paola},
	year = {2022},
}

@article{forsberg2014diffusion,
  title={Diffusion in the study of civil wars: A cautionary tale},
  author={Forsberg, Erika},
  journal={International Studies Review},
  volume={16},
  number={2},
  pages={188--198},
  year={2014},
  publisher={Blackwell Publishing Ltd Oxford, UK}
}

@article{saideman2012conflict,
  title={When conflict spreads: Arab spring and the limits of diffusion},
  author={Saideman, Stephen M},
  journal={International Interactions},
  volume={38},
  number={5},
  pages={713--722},
  year={2012},
  publisher={Taylor \& Francis}
}

@article{kibris2021geo,
  title={The geo-temporal evolution of violence in civil conflicts: A micro analysis of conflict diffusion on a new event data set},
  author={Kibris, Arzu},
  journal={Journal of Peace Research},
  volume={58},
  number={5},
  pages={885--899},
  year={2021},
  publisher={SAGE Publications Sage UK: London, England}
}

\setcounter{figure}{0}
\clearpage
\goodbreak
\appendix
\section{Example of forecast : Prio-Grid level} \label{app:prio}

\begin{figure}[h!]
    \centering
    \begin{subfigure}{\textwidth}
        \centering
        \includegraphics[width=\textwidth]{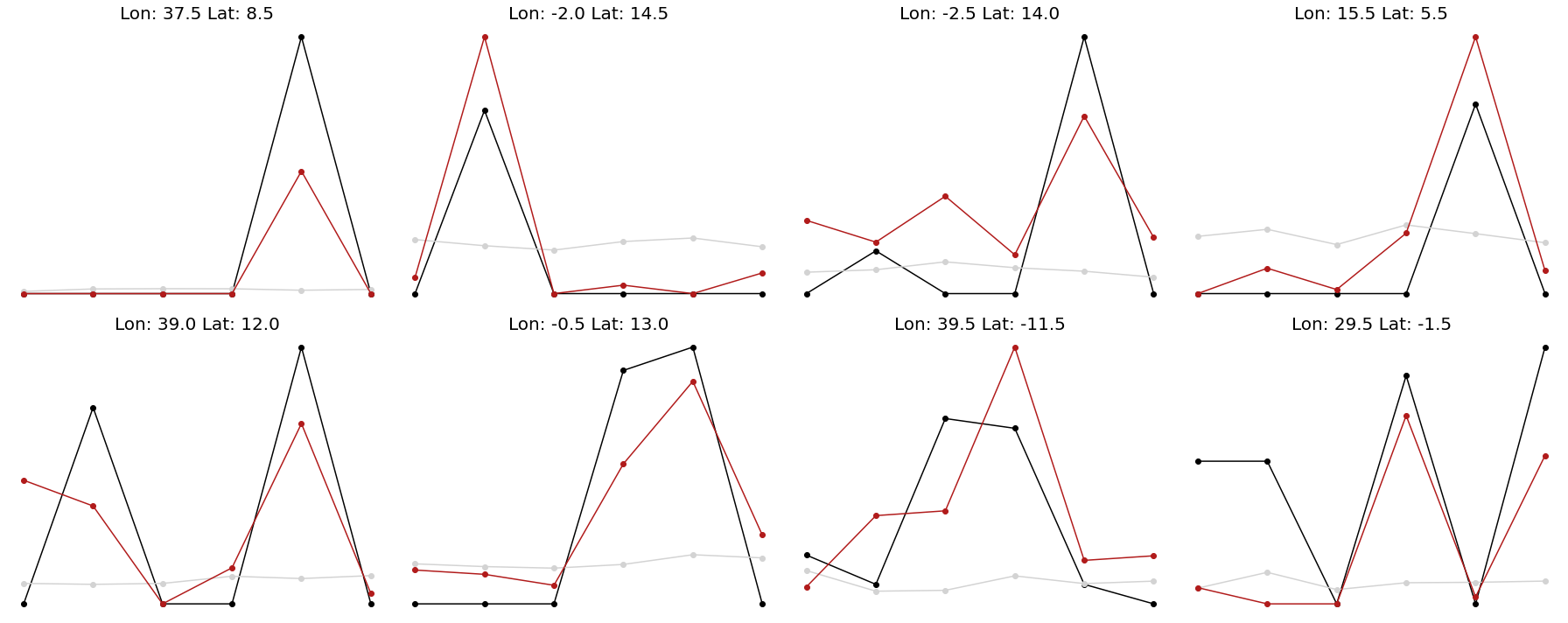}
        \subcaption{Best performing cases}
    \end{subfigure}
    \begin{subfigure}{\textwidth}
        \centering
        \includegraphics[width=\textwidth]{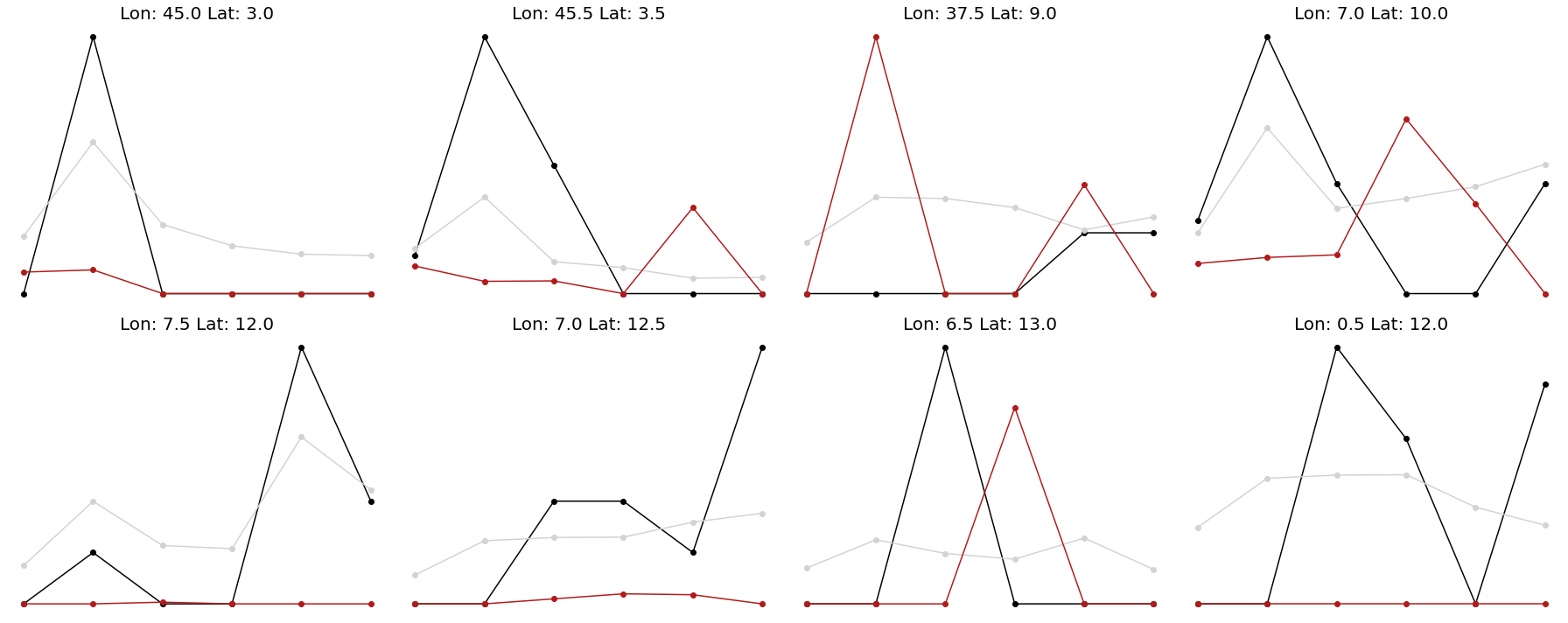}
        \subcaption{Worse performing cases}
    \end{subfigure}
    \caption{Examples of best (top) and worst (bottom) Prio-Grid MSE scores. The red line is the SF values, the grey is Views, and the black is the observed values.}
    \label{app:ex_prio}
\end{figure}

\setcounter{figure}{0}
\clearpage
\section{Example of forecast : Pattern level} \label{app:pattern}

\begin{figure}[h!]
    \centering
    \begin{subfigure}{\textwidth}
        \centering
        \includegraphics[width=0.8\textwidth]{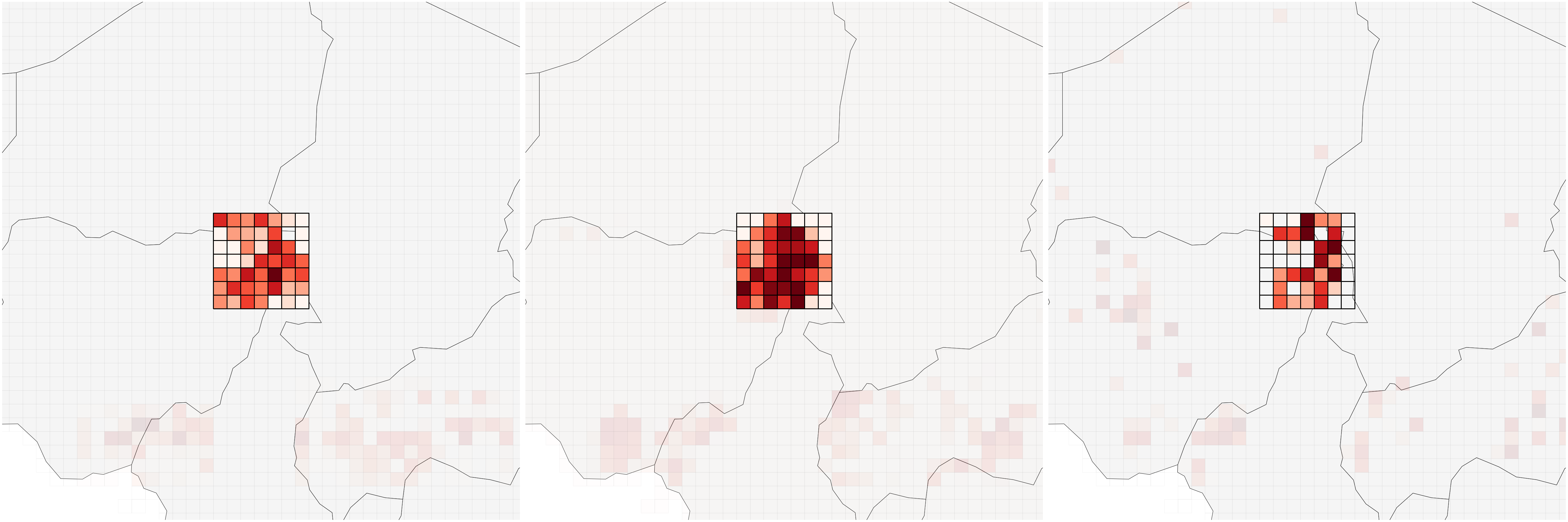}
        \subcaption{Lake Chad area - Jan-June 2022}
    \end{subfigure}
    \begin{subfigure}{\textwidth}
        \centering
        \includegraphics[width=0.8\textwidth]{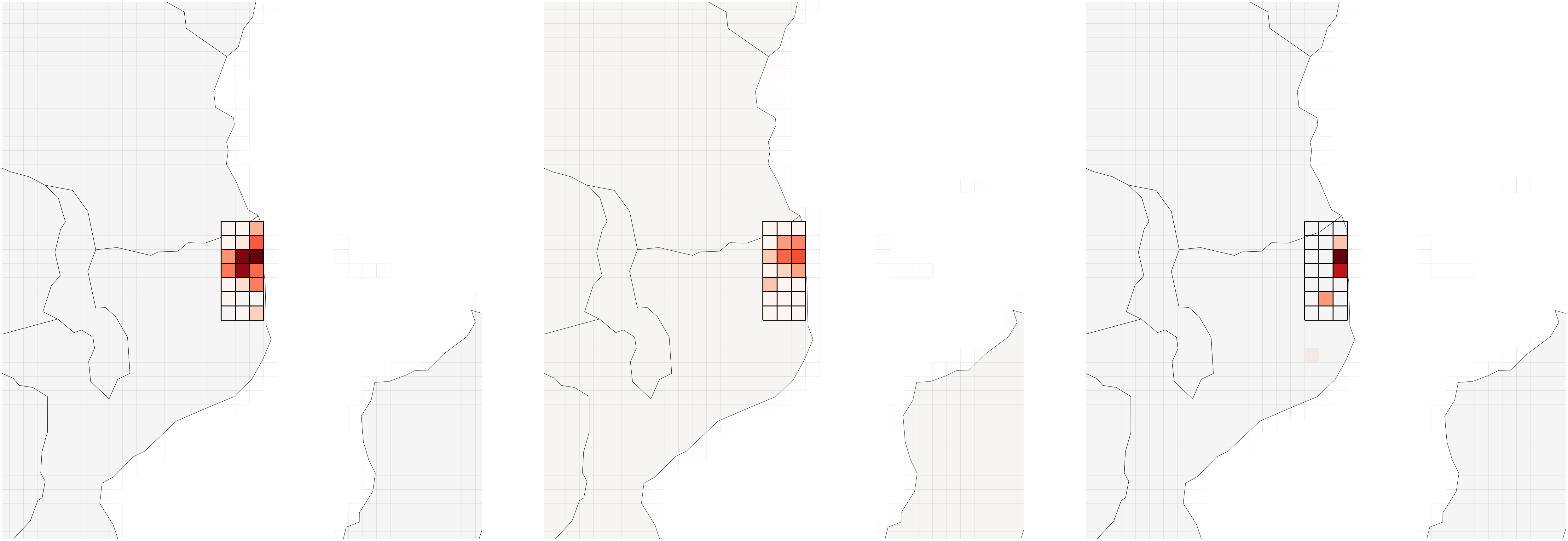}
        \subcaption{North Mozambique - July-Dec 2023}
    \end{subfigure}
    \begin{subfigure}{\textwidth}
        \centering
        \includegraphics[width=0.8\textwidth]{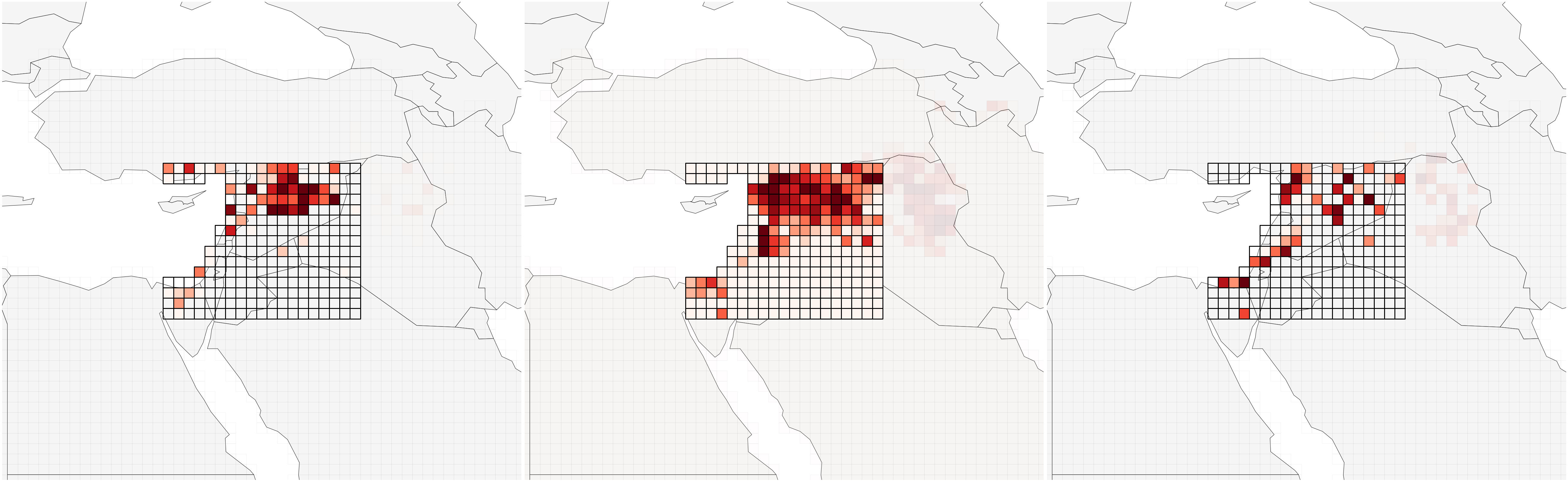}
        \subcaption{Levant Region - Jan-June 2022}
    \end{subfigure}
    \begin{subfigure}{\textwidth}
        \centering
        \includegraphics[width=0.8\textwidth]{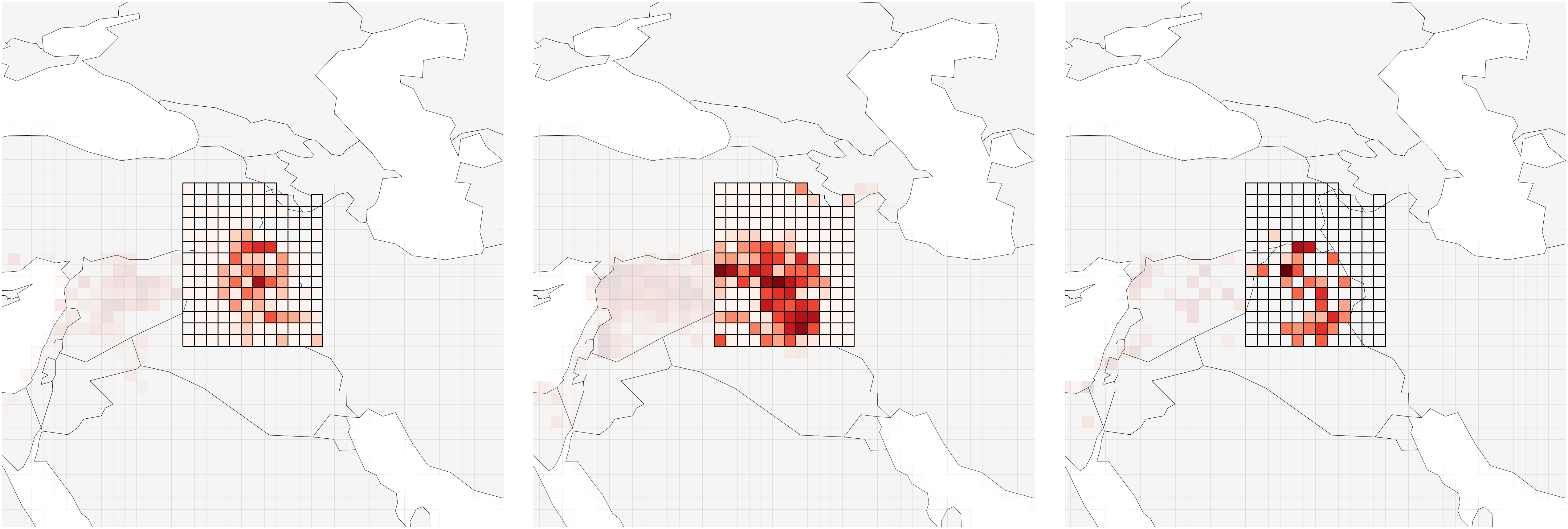}
        \subcaption{North Iraq - Jan-June 2022}
    \end{subfigure}
    \caption{Examples of four Pattern ShapeFinder Forecast (left), Views Forecast (middle), and Observed values (right). The darker the red, the higher the fatality value. The values are aggregated per Prio-Grid over the 6-month period.}
    \label{app:ex_patterns}
\end{figure}

\setcounter{figure}{0}
\clearpage
\section{Under/over prediction robustness test} \label{app:over}

\begin{figure}[h]
\centering
\includegraphics[width=0.7\textwidth]{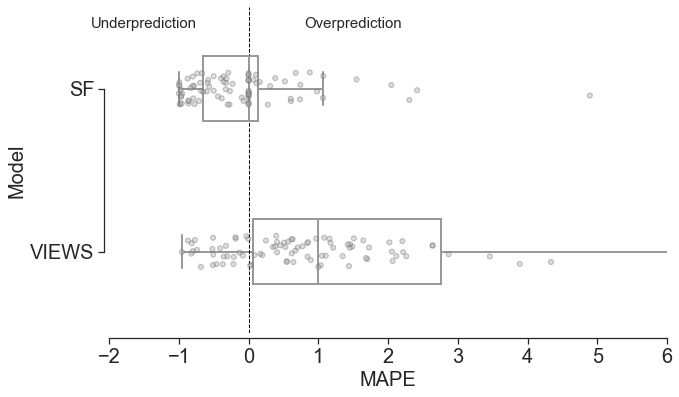}  \caption{Boxplot of the Mean Average Percentage Error (MAPE) for both models, SF and Views. A positive value corresponds to an overprediction, in contrast to a negative value.}
\label{app:glob_over}
\end{figure}

\begin{figure}[h]
\centering
\includegraphics[width=0.7\textwidth]{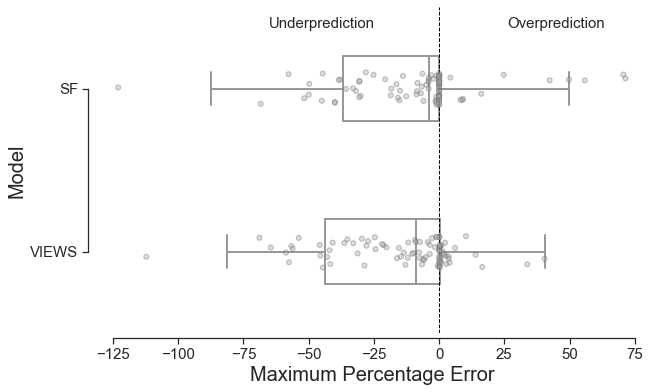}  \caption{Boxplot of the Maximum Percentage Error for both models, SF and Views. A positive value corresponds to an overprediction, in contrast to a negative value.}
\label{app:max_over}
\end{figure}

\setcounter{figure}{0}
\clearpage
\section{Kenya-Somalia example} 
\label{app:projo}

\begin{figure}[h]
    \centering
    \includegraphics[width=\textwidth]{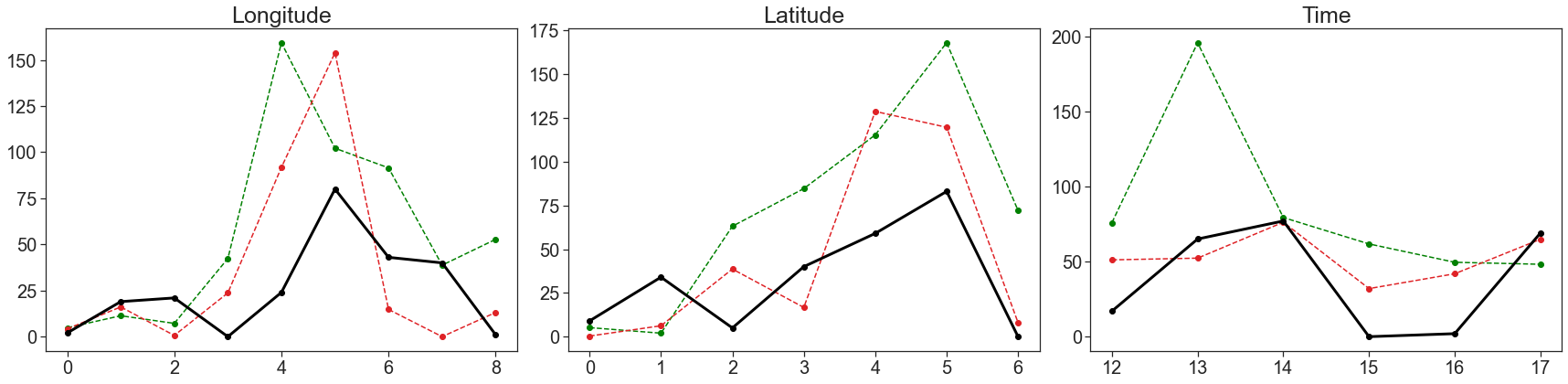}
    \caption{Projection of fatalities values over the Longitude (left), Latitude (middle), and Time (right). The black line corresponds to the observed values, the green to the Views prediction, and the red to the ShapeFinder prediction.}
    \label{projection}
\end{figure}

Figure \ref{projection} compares the forecasted projections of fatalities from the ShapeFinder (SF) model and the Views model against the actual observed values. The black line represents the observed data, while the green and red lines correspond to the Views and ShapeFinder predictions. Our model demonstrates a clear ability to capture the peak of fatalities accurately in both longitude and time dimensions. However, it misses the peak value in latitude by one unit, indicating a minor spatial misalignment. In contrast, the Views model exhibits a consistent overprediction, especially in the first two-month horizon.  

\setcounter{figure}{0}
\clearpage
\section{Diffusion of fatalities to neighboring cells} 

\begin{figure}[h]
    \centering
    \includegraphics[width=\textwidth]{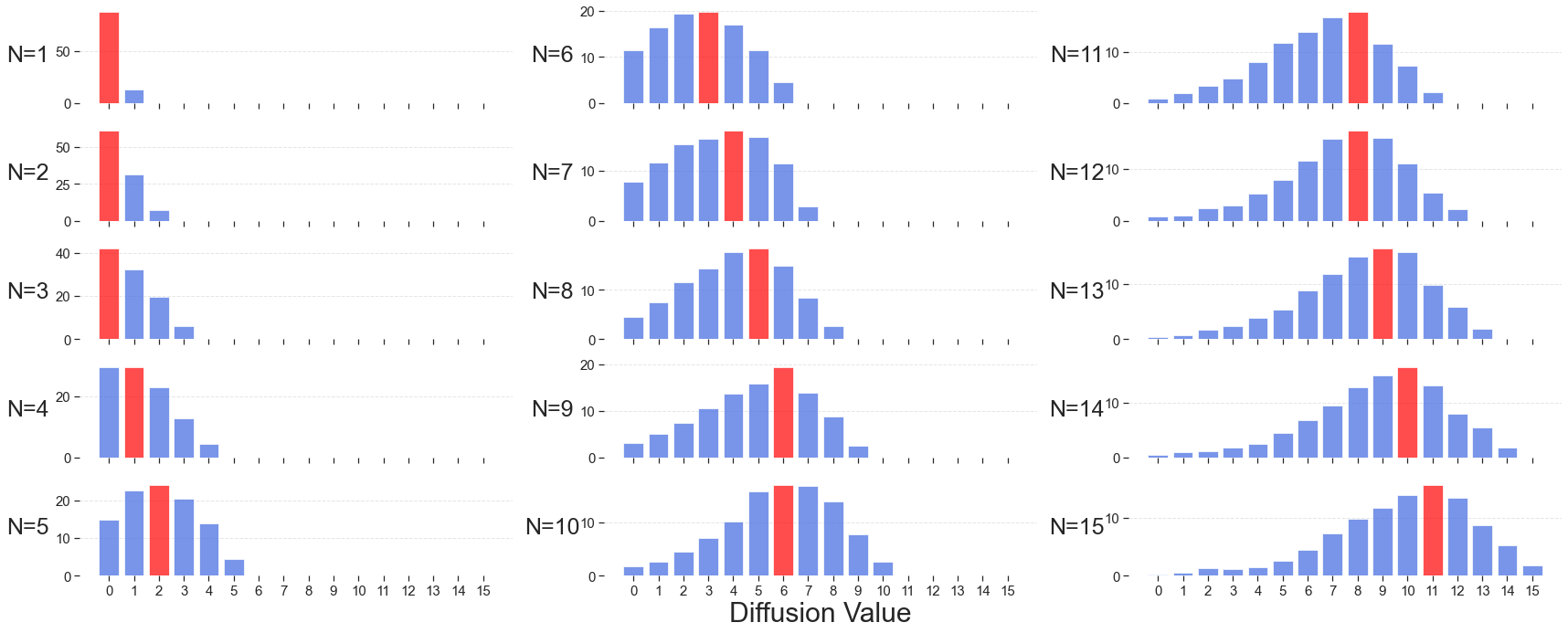}
    \caption{Histogram of the number of active cells in the spatial neighborhood (5×5 grid) during the future window (t+1 to t+6), conditioned on the number of active neighboring cells during the past window (t-12 to t). Each panel corresponds to a different level of past activity (N = 1 to 15), where N is the number of active cells in the 5×5 neighborhood from t-12 to t. The red bar in each histogram indicates the most frequent outcome.}
    \label{app:projo}
\end{figure}

\clearpage
\section{Additional Results: Test Periods and Forecasting Horizon}
\label{app:add_res}

\begin{table}[htbp]
\centering
\caption{Absolute Error Log Ratio: $\log\left(\frac{\text{AE}_{\text{Views}} + 1}{\text{AE}_{\text{SF}} + 1}\right)$}
\label{tab:mae_log_ratio}
\begin{tabular}{lcccccccc}
\hline
Period & $h=1$ & $h=2$ & $h=3$ & $h=4$ & $h=5$ & $h=6$ & All Horizons \\
\hline
Jan-2022 & 0.2287 & 0.3092 & 0.3638 & 0.3547 & 0.3550 & 0.3391 & 0.3251 \\
         & (0.0191) & (0.0219) & (0.0230) & (0.0244) & (0.0235) & (0.0238) & (0.0093) \\
July-2022 & 0.1742 & 0.2397 & 0.1929 & 0.2135 & 0.1941 & 0.1920 & 0.2011 \\
          & (0.0177) & (0.0191) & (0.0184) & (0.0181) & (0.0175) & (0.0179) & (0.0074) \\
Jan-2023 & 0.0994 & 0.2289 & 0.1114 & 0.1731 & 0.1632 & 0.1444 & 0.1534 \\
         & (0.0176) & (0.0206) & (0.0188) & (0.0176) & (0.0174) & (0.0186) & (0.0075) \\
July-2023 & 0.1174 & 0.1851 & 0.1829 & 0.2164 & 0.2400 & 0.3004 & 0.2070 \\
          & (0.0155) & (0.0157) & (0.0142) & (0.0151) & (0.0168) & (0.0177) & (0.0065) \\
\hline
All Periods & 0.1534 & 0.2391 & 0.2107 & 0.2379 & 0.2371 & 0.2441 & 0.2204 \\
            & (0.0087) & (0.0097) & (0.0094) & (0.0095) & (0.0095) & (0.0098) & (0.0039) \\
\hline
\end{tabular}
\begin{tablenotes}
\small
\item Note: Standard errors in parentheses. Positive values indicate the ShapeFinder (SF) model outperforms the Views model. 
\end{tablenotes}
\end{table}

\begin{table}[htbp]
\centering
\caption{Squared Error (SE) Log Ratio: $\log\left(\frac{\text{SE}_{\text{Views}} + 1}{\text{SE}_{\text{SF}} + 1}\right)$}
\label{tab:mse_log_ratio}
\begin{tabular}{lcccccccc}
\hline
Period & $h=1$ & $h=2$ & $h=3$ & $h=4$ & $h=5$ & $h=6$ & All Horizons \\
\hline
Jan-2022 & 0.3117 & 0.4574 & 0.5650 & 0.5533 & 0.5464 & 0.5197 & 0.4923 \\
         & (0.0374) & (0.0436) & (0.0458) & (0.0487) & (0.0471) & (0.0470) & (0.0184) \\
July-2022 & 0.1762 & 0.2975 & 0.2155 & 0.2367 & 0.2043 & 0.2053 & 0.2226 \\
          & (0.0333) & (0.0361) & (0.0352) & (0.0341) & (0.0328) & (0.0334) & (0.0140) \\
Jan-2023 & 0.0604 & 0.3159 & 0.0625 & 0.1699 & 0.1630 & 0.1116 & 0.1472 \\
         & (0.0332) & (0.0403) & (0.0360) & (0.0337) & (0.0333) & (0.0357) & (0.0145) \\
July-2023 & 0.0931 & 0.1983 & 0.1903 & 0.2358 & 0.2816 & 0.3941 & 0.2322 \\
          & (0.0289) & (0.0295) & (0.0259) & (0.0284) & (0.0319) & (0.0340) & (0.0122) \\

\hline
All Periods & 0.1574 & 0.3139 & 0.2540 & 0.2955 & 0.2963 & 0.3073 & 0.2708 \\
            & (0.0166) & (0.0187) & (0.0181) & (0.0184) & (0.0183) & (0.0189) & (0.0074) \\
\hline
\end{tabular}
\begin{tablenotes}
\small
\item Note: Standard errors in parentheses. Positive values indicate the ShapeFinder (SF) model outperforms the Views model. 
\end{tablenotes}
\end{table}

Tables \ref{tab:mae_log_ratio} and \ref{tab:mse_log_ratio} report the log differences in Mean Absolute Error (MAE) and Mean Squared Error (MSE) between the Views model and ShapeFinder. Positive values indicate that ShapeFinder outperforms the Views model. Values in parentheses are standard errors, calculated by dividing the standard deviation of the subgroup observations by the square root of the number of observations. The periods Jan-2022, Jul-2022, Jan-2023, and Jul-2023 include 28, 26, 27, and 30 active conflict zones, corresponding to 1,505, 1,535, 1,577, and 1,676 Prio-Grid cells, respectively.

The first notable result is the consistency of performance across both forecast horizons and test periods. Looking at the MAE results in Table \ref{tab:mae_log_ratio}, the model consistently outperforms across all periods and horizons, with log ratios ranging from 0.0994 to 0.3638. The Jan-2022 period has the strongest performance with an overall log ratio of 0.3251 (SE: 0.0093), indicating lower forecast errors compared to the Views model. Performance peaks at horizon 3 for this period (0.3638), while horizon 1 shows more modest improvements across all periods. The July-2023 period shows an interesting pattern where performance improves monotonically with forecast horizon, reaching 0.3004 at $h=6$. Across all periods and horizons, the aggregate log ratio of 0.2204 (SE: 0.0039) confirms systematic improvement over the Views model.

The MSE results in Table \ref{tab:mse_log_ratio} reveal a similar pattern. The Jan-2022 period again demonstrates the strongest performance with an overall log ratio of 0.4923 (SE: 0.0184), nearly double that of the Jan-2023 period (0.1472). Notably, horizon 2 consistently shows strong performance across all periods, with values ranging from 0.1983 to 0.4574. The Jan-2022 period indicates remarkably stable performance from $h=3$ to $h=6$ (all above 0.51).

\clearpage
\setcounter{figure}{0}
\section{Concepts Definitions}
\label{app:def}

\begin{definition}[Sequence]
Let $\{x_{i,j,t}\}$ be a Prio-Grid monthly fatality field where $(i,j)$ indexes a grid cell and $t \in \{1,\dots,T\}$ indexes discrete monthly time steps.

A \emph{sequence} of temporal length $L$ months and spatial extent $(W_{\text{lat}}, W_{\text{lon}})$ grid cells is a spatiotemporal segment defined as
\[
S_{i_0,j_0,t_0;W_{\text{lat}},W_{\text{lon}},L} = \{x_{i,j,t} : i_0 \leq i \leq i_0+W_{\text{lat}}-1, \, j_0 \leq j \leq j_0+W_{\text{lon}}-1, \, t_0 \leq t \leq t_0+L-1\}.
\]
\end{definition}

\vspace{1cm}

\begin{definition}[Past Future]
Given a spatiotemporal sequence $S_{i_0,j_0,t_0;W_{\text{lat}},W_{\text{lon}},L}$, its \emph{Past Future} is the set of fatality values in the same spatial region over the next $H$ months:
\[
PF_{i_0,j_0,t_0;W_{\text{lat}},W_{\text{lon}},H} = \{x_{i,j,t} : i_0 \leq i \leq i_0+W_{\text{lat}}-1, \, j_0 \leq j \leq j_0+W_{\text{lon}}-1, \, t_0+L \leq t \leq t_0+L+H-1\}.
\]
$H$ in this paper is set to six months.
\end{definition}
\vspace{1cm}

\begin{definition}[Pattern]
A pattern is a recurring three-dimensional shape that may have several occurrences in history. In the  ShapeFinder model, an input sequence and the historical similar sequences are occurrences of the same pattern.
\par Formally, a pattern is the set of sequences whose EMD distance (Equation 1) and R ratio (Equation 2) are below thresholds $thr_{1}$ and $thr_{2}$.
\end{definition}
\vspace{1cm}

\begin{definition}[Active cell]
An active cell is a Prio-Grid cell that has more than zero fatalities for a given period of time. In Figure \ref{ex_pattern_1}, the active cells are colored in red, for the year 2022. 
\end{definition}
\vspace{1cm}

\begin{definition}[Active zone]
An active zone is an aggregation of minimum two active cells, within a radius of 2 spatial unit. In a zone, each active cell is maximum distant to 2 unit from another active cell. Figure \ref{ex_pattern_4} shows examples of active zones in 2022.   
\end{definition}
\vspace{1cm}

\setcounter{figure}{0}
\clearpage
\section{Impact of the active zone selection}
\label{imp_acs}

Even if the zoning selection covers almost the entire short-term autoregressive information, in this section, the coverage of the test zone is assessed through two measures: the percentage of fatalities in the test set covered by the zone defined by the training set and the percentage of active cells. Then, the impact on the prediction error is assessed, as the Views model covers the entire zone without distinction. In other words, the validity of a pure autoregressive model is assessed in the conflict fatalities setup. 

\begin{figure}[!h]
    \centering
    \includegraphics[width=0.7\textwidth]{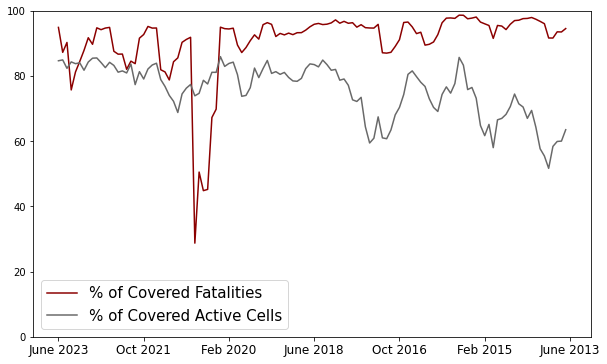}
    \caption{Time series of the monthly percentage of inclusion of fatalities (in blue) and active cells (in orange) in the active zones. The y-axis represents the percentage, and the x-axis represents the last month included in the training set used to define the active zones. E.g., the values indicated at x=June 2015, show the results of the test with a training set from July 2014 to June 2015, and the test set from August 2015 to January 2016.}
    \label{ts_perc}
\end{figure}

In this test, we simulate the selection of active zones, as illustrated in Figure \ref{ex_pattern}, using a rolling training window from July 2012–June 2013 through July 2022–June 2023. For each window, the active zone method is applied, and we measure the share of active cells and fatalities occurring within these zones during the following six months. In the first iteration, fatalities aggregated over July 2012–June 2013 are used to define the active zones. We then calculate the percentage of fatalities and active cells from July to December 2013 that fall within these zones. The window is subsequently shifted forward by one month, with August 2012–July 2013 used as the next training period, and the process is repeated. This final result is a time series of monthly inclusion rates for fatalities and active cells within the active zones. The series is plotted in Figure \ref{ts_perc}.

On average, the active zones capture more than 90\% of fatalities and about 75\% of active cells. This suggests that the method performs well in estimating the broad spatial location of future conflict. However, still around 10\% of fatalities and 25\% of active cells fall outside the predicted zones. These missed cases mostly correspond to the onset of conflict, which are often low in intensity and harder to detect, explaining the difference between the two measures. The period in late 2020 shows a clear limit of the model. It represents the onset of the Tigray war in the north of Ethiopia, in a post-Covid period with very few fatalities, producing inflated percentage values. 

\begin{figure}[!h]
    \centering
    \includegraphics[width=0.7\textwidth]{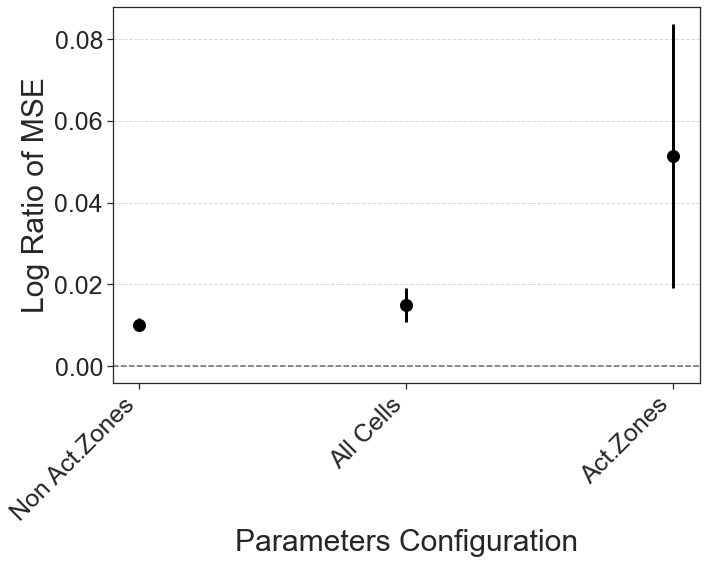}
    \caption{Average Log ratio of MSE for the Prio-Grid period observations, comparing the Views model and the ShapeFinder (SF). The black point is the average value, and the 95\% confidence interval is calculated with the standard error divided by the square root of the number of observations. When the ratio is positive, the SF model outperforms the Views model. The results are presented for three categories: The cells not included in the active zones (left), all the cells (center), and the cells in the active zones (right).}
    \label{zero_test}
\end{figure}

To assess how this limitation affects model performance, Figure \ref{zero_test} shows the logged difference in mean squared error between the Views model and ShapeFinder (SF). The black dot represents the average difference, and the 95\% confidence interval is computed using the standard error divided by the square root of the number of observations. A positive value indicates that ShapeFinder outperforms the Views model. Results are reported for three groups of cells: those outside the active zones (left), all cells (center), and those within the active zones (right). Across all three groups, ShapeFinder consistently outperforms the Views model. The difference in error is largest for cells within active zones, where accurate predictions matter most. Outside the active zones, the difference in error narrows and approaches zero, indicating more similar performance across models. In these areas, ShapeFinder only predicts zero fatalities, indicating that the Views also model makes conservative forecasts with few fatalities. 

\clearpage
\setcounter{figure}{0}
\section{Comparison with the original ShapeFinder}
\label{app:comp_SF}

As a robustness test of the three-dimensional spatio-temporal adaptation, we compare its forecasting performance with that of the original ShapeFinder model. The original model was designed for country-month data. In this comparison, each Prio-Grid cell is treated as an independent country. The input consists of the previous 12 months of fatalities for each cell, and the full set of Prio-Grid cell fatality time series serves as the historical reference. 

\begin{figure}[!h]
    \centering
    \includegraphics[width=0.8\textwidth]{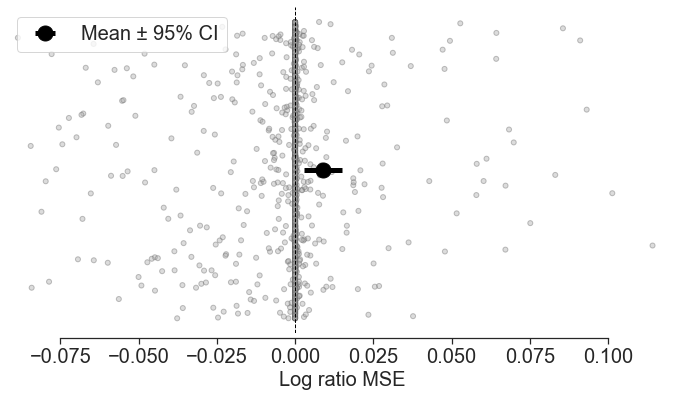}
    \caption{Log ratio in Mean Squared Error of the original ShapeFinder model and the three-dimensional adaptation. A positive value corresponds to a lower error from the adaptation, in contrast to a negative value that corresponds to a better performance from the original model. The grey points represent the ratio for one Prio-Grid cell, while the black point and line represent the average value with the 95\% confidence interval, calculated using the standard error divided by the square root of the number of observations.}
    \label{comp_sf_ori}
\end{figure}

Because the historical dataset is very large, the test is limited to an input period from January 2022 to December 2022, with forecasts evaluated for the subsequent six months, from January 2023 to June 2023, using historical data available up to December 2022. The original ShapeFinder model requires about one hour to run for roughly 120 countries. In this grid-cell application, even after filtering out the least active cells, the model took more than one day to complete. Performance results are shown in Figure \ref{comp_sf_ori}. Overall, the three-dimensional model consistently outperforms the original ShapeFinder in accuracy. Given its lower performance and substantially higher runtime, the original model is not suitable for large-scale grid-level forecasting. However, it may still be useful for isolated cells where the three-dimensional model cannot be applied.

\clearpage
\setcounter{figure}{0}
\section{Grid example calculation}
\label{app:calcu}

\begin{figure}[h]
    \centering
    \includegraphics[width=0.9\textwidth]{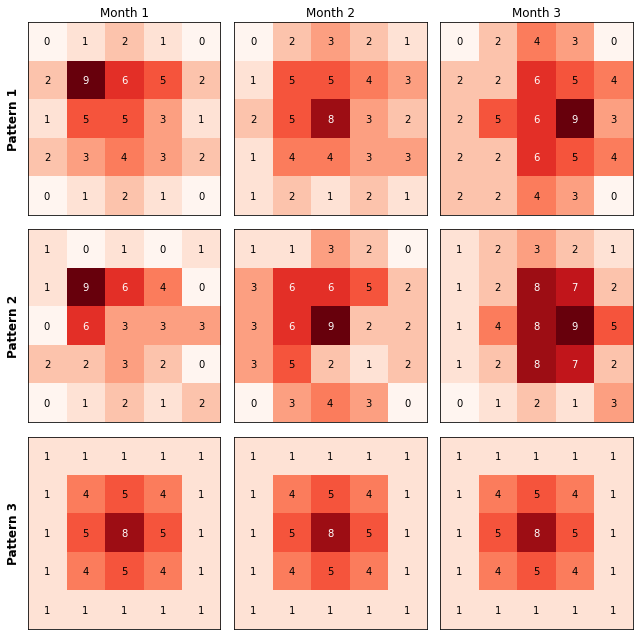}
    \caption{Representation of the example shown in Figure \ref{toy}. Three patterns on a 5×5 grid over three months. The x-axis represents longitude, and the y-axis represents latitude. The cell color indicates the number of fatalities per month, ranging from 0 (white) to 9 (dark red), with darker shades representing higher values. The number in the center of the cell indicates the value of the cell.}
    \label{toy_number}
\end{figure}

In this section, the calculations of the Spatial Distance,  Earth Mover's Distance (EMD), and Euclidean distance of the example presented in Figure \ref{toy} are detailed. As a reminder, the patterns are represented in Figure \ref{toy_number} with the cell values. 

\clearpage
\setcounter{figure}{0}
\subsection*{Spatial Distance}

In classical spatial regressions, the model includes a spatial lag, with higher weights assigned to the closest neighbors. In our example, where the central cell is the point of interest, the weights would be:
\[
weights=
\begin{bmatrix}
0.01 & 0.05 & 0.1  & 0.05 & 0.01 \\
0.05 & 0.3  & 0.5  & 0.3  & 0.05 \\
0.1  & 0.5  & 1.0  & 0.5  & 0.1  \\
0.05 & 0.3  & 0.5  & 0.3  & 0.05 \\
0.01 & 0.05 & 0.1  & 0.05 & 0.01
\end{bmatrix}
\]

In the matrix, the highest weight is placed at the center (cell [2,2] in the matrix), and the weights decrease as the distance from the central cell increases. These weights are proportional to the distance from the center of the grid.  

Then, if we define the spatial distance $SP_{P_{1},P_{2}}$ between pattern 1 and pattern 2 with:
\[
SP_{P_{1},P_{2}}= \sum_{month} |\sum_{lat,long}P_{1,lat,long,m}*weights_{lat,long} - \sum_{lat,long}P_{2,lat,long,m}*weights_{lat,long}|
\]

Then :
\begin{align*}
\alpha &= P_{1,lat,long,m}*weights_{lat,long} \\
&=
\begin{bmatrix}
\begin{bmatrix}
0.00 & 0.05 & 0.20 & 0.05 & 0.00 \\
0.10 & 2.70 & 3.00 & 1.50 & 0.10 \\
0.10 & 2.50 & 5.00 & 1.50 & 0.10 \\
0.10 & 0.90 & 2.00 & 0.90 & 0.10 \\
0.00 & 0.05 & 0.20 & 0.05 & 0.00
\end{bmatrix},
\begin{bmatrix}
0.00 & 0.10 & 0.30 & 0.10 & 0.01 \\
0.05 & 1.50 & 2.50 & 1.20 & 0.15 \\
0.20 & 2.50 & 8.00 & 1.50 & 0.20 \\
0.05 & 1.20 & 2.00 & 0.90 & 0.15 \\
0.01 & 0.10 & 0.10 & 0.10 & 0.01
\end{bmatrix},
\begin{bmatrix}
0.00 & 0.10 & 0.40 & 0.15 & 0.00 \\
0.10 & 0.60 & 3.00 & 1.50 & 0.20 \\
0.20 & 2.50 & 6.00 & 4.50 & 0.30 \\
0.10 & 0.60 & 3.00 & 1.50 & 0.20 \\
0.02 & 0.10 & 0.40 & 0.15 & 0.00
\end{bmatrix}
\end{bmatrix}
\end{align*}

\begin{align*}
\beta &= P_{2,lat,long,m}*weights_{lat,long} \\
&=
\begin{bmatrix}
\begin{bmatrix}
0.01 & 0.00 & 0.10 & 0.00 & 0.01 \\
0.05 & 2.70 & 3.00 & 1.20 & 0.00 \\
0.00 & 3.00 & 3.00 & 1.50 & 0.30 \\
0.10 & 0.60 & 1.50 & 0.60 & 0.00 \\
0.00 & 0.05 & 0.20 & 0.05 & 0.02
\end{bmatrix},
\begin{bmatrix}
0.01 & 0.05 & 0.30 & 0.10 & 0.00 \\
0.15 & 1.80 & 3.00 & 1.50 & 0.10 \\
0.30 & 3.00 & 9.00 & 1.00 & 0.20 \\
0.15 & 1.50 & 1.00 & 0.30 & 0.10 \\
0.00 & 0.15 & 0.40 & 0.15 & 0.00
\end{bmatrix},
\begin{bmatrix}
0.01 & 0.10 & 0.30 & 0.10 & 0.01 \\
0.05 & 0.60 & 4.00 & 2.10 & 0.10 \\
0.10 & 2.00 & 8.00 & 4.50 & 0.50 \\
0.05 & 0.60 & 4.00 & 2.10 & 0.10 \\
0.00 & 0.05 & 0.20 & 0.05 & 0.03
\end{bmatrix}
\end{bmatrix}
\end{align*}

So:
\begin{align*}
SP_{P_{1},P_{2}} &= \sum_{month}|\sum \alpha_{month}-\sum \beta_{month}| \\
&= 3.21+ 1.33 + 4.03 = 8.57
\end{align*}

Similarly for $SP_{P_{1},P_{3}}$,
\begin{align*}
\delta &= P_{3,lat,long,m}*weights_{lat,long} \\
&=
\begin{bmatrix}
\begin{bmatrix}
0.01 & 0.05 & 0.10 & 0.05 & 0.01 \\
0.05 & 1.20 & 2.50 & 1.20 & 0.05 \\
0.10 & 2.50 & 8.00 & 2.50 & 0.10 \\
0.05 & 1.20 & 2.50 & 1.20 & 0.05 \\
0.01 & 0.05 & 0.10 & 0.05 & 0.01
\end{bmatrix},
\begin{bmatrix}
0.01 & 0.05 & 0.10 & 0.05 & 0.01 \\
0.05 & 1.20 & 2.50 & 1.20 & 0.05 \\
0.10 & 2.50 & 8.00 & 2.50 & 0.10 \\
0.05 & 1.20 & 2.50 & 1.20 & 0.05 \\
0.01 & 0.05 & 0.10 & 0.05 & 0.01
\end{bmatrix},
\begin{bmatrix}
0.01 & 0.05 & 0.10 & 0.05 & 0.01 \\
0.05 & 1.20 & 2.50 & 1.20 & 0.05 \\
0.10 & 2.50 & 8.00 & 2.50 & 0.10 \\
0.05 & 1.20 & 2.50 & 1.20 & 0.05 \\
0.01 & 0.05 & 0.10 & 0.05 & 0.01
\end{bmatrix}
\end{bmatrix}
\end{align*}

So:
\begin{align*}
SP_{P_{1},P_{3}} &= \sum_{month}|\sum \alpha_{month}-\sum \delta_{month}| \\
&= 2.44+0.71+ 1.98 = 5.13
\end{align*}

\clearpage
\section*{Earth Mover's Distance}

The Earth Mover's Distance (EMD) is computed as follows:
\begin{equation*}
\text{EMD} = \sum_{i,j} \gamma_{i,j}^{*} * c_{i,j}
\end{equation*}
where $\gamma_{i,j}^{*}$ is the Optimal Transport Matrix or the matrix that lowers the transport cost. $c_{i,j}$ is the cost matrix of the Euclidean distance between the points $i$ and $j$. 

The Optimal Transport Matrix is computed from the ``Source'' distribution called $w_{1}$ (i.e., the normalized fatalities of pattern 1) and the ``Target'' distribution $w_{2}$ (i.e., the normalized fatalities of pattern 2). 
 
The normalized patterns are computed as:
\begin{equation*}
P_1^{\text{norm}} = \frac{P_1}{\sum P_1}, \quad
P_2^{\text{norm}} = \frac{P_2}{\sum P_2}
\end{equation*}

This gives us the following points : 
\begin{table}[h]
    \centering
    \begin{tabular}{ccc|c||ccc|c}
        \multicolumn{4}{c}{\textbf{Pattern 1}} & \multicolumn{4}{c}{\textbf{Pattern 2}} \\
        \toprule
        Longitude & Latitude & Month &$Fatatilies_{norm}$ & Longitude & Latitude & Month &$Fatatilies_{norm}$\\
        \midrule
        0.0000 & 0.0000 & 1.0000 & 0.0047 & 0.0000 & 0.0000 & 0.0000 & 0.0048 \\
        0.0000 & 0.0000 & 2.0000 & 0.0094 & 0.0000 & 0.0000 & 2.0000 & 0.0048 \\
        0.0000 & 0.0000 & 3.0000 & 0.0047 & 0.0000 & 0.0000 & 4.0000 & 0.0048 \\
        0.0000 & 1.0000 & 0.0000 & 0.0094 & 0.0000 & 1.0000 & 0.0000 & 0.0048 \\
        0.0000 & 1.0000 & 1.0000 & 0.0425 & 0.0000 & 1.0000 & 1.0000 & 0.0429 \\
        0.0000 & 1.0000 & 2.0000 & 0.0283 & 0.0000 & 1.0000 & 2.0000 & 0.0286 \\
        \vdots & \vdots & \vdots & \vdots & \vdots & \vdots & \vdots & \vdots\\
        2.0 & 4.0 & 2.0 & 0.0189 & 2.0 & 4.0 & 3.0 & 0.0048\\
        2.0 & 4.0 & 3.0 & 0.0142 & 2.0 & 4.0 & 4.0 & 0.0143\\
        \bottomrule
    \end{tabular}
\end{table}

After computation of the optimal transport matrix, we obtain : 
\begin{equation*}
\gamma_{i,j}^{*} = 
\begin{bmatrix}
0.0046 & 0.0 & 0.0 & ... & 0.0 & 0.0 & 0.0 \\
0.0 & 0.0048 & 0.0 & ... & 0.0 & 0.0 & 0.0 \\
0.0 & 0.0 & 0.0046 & ... & 0.0 & 0.0 & 0.0 \\
\vdots & \vdots & \vdots & \ddots & \vdots & \vdots & \vdots \\
0.0 & 0.0 & 0.0 & ... & 0.0 & 0.0 & 0.0 \\
0.0 & 0.0 & 0.0 & ... & 0.0095 & 0.0 & 0.0 \\
0.0 & 0.0 & 0.0 & ... & 0.0 & 0.0048 & 0.009 \\
\end{bmatrix}
\end{equation*}

Then, the Euclidean distance between the coordinates of both patterns is calculated using:
\begin{equation*}
d_{i,j} = \sqrt{(long_{i}-long_{j})^2+(lat_{i}-lat_{j})^2+(month_{i}-month_{j})^2}
\end{equation*}

This results in a cost matrix $c_{i,j}$ of shape $(n_{1},n_{2})$, where $n_{1}$ is the number of non-zero cell in pattern1, and $n_{2}$ the number of non-zero cell in pattern2. 

\begin{equation*}
c_{i,j} = 
\begin{bmatrix}
1.0 & 1.0 & 3.0 & ... & 4.5826 & 4.899 & 5.3852 \\
2.0 & 0.0 & 2.0 & ... & 4.4721 & 4.5826 & 4.899 \\
3.0 & 1.0 & 1.0 & ... & 4.5826 & 4.4721 & 4.5826 \\
\vdots & \vdots & \vdots & \ddots & \vdots & \vdots & \vdots \\
4.5826 & 4.5826 & 5.3852 & ... & 1.0 & 2.0 & 3.0 \\
4.899 & 4.4721 & 4.899 & ... & 0.0 & 1.0 & 2.0 \\
5.3852 & 4.5826 & 4.5826 & ... & 1.0 & 0.0 & 1.0 \\
\end{bmatrix}
\end{equation*}

So: 
\begin{align*}
    EMD_{P_{1},P_{2}} &= \sum_{i,j} \gamma_{i,j}^{*} * c_{i,j} \\
    &= (0.0046 \times 1.0) + (0.0046 \times 1.0) + ... + (0.009 \times 1.0) \\
        &= 0.2290
\end{align*}

Similarly, we repeat the operation with $P_{1}$ and $P_{3}$. 

\begin{equation*}
\gamma_{i,k}^{*} = 
\begin{bmatrix}
0.0 & 0.0047 & 0.0 & ... & 0.0 & 0.0 & 0.0 \\
0.0 & 0.0008 & 0.0056 & ... & 0.0 & 0.0 & 0.0 \\
0.0 & 0.0 & 0.0 & ... & 0.0 & 0.0 & 0.0 \\
\vdots & \vdots & \vdots & \ddots & \vdots & \vdots & \vdots \\
0.0 & 0.0 & 0.0 & ... & 0.0 & 0.0 & 0.0 \\
0.0 & 0.0 & 0.0 & ... & 0.0056 & 0.0 & 0.0 \\
0.0 & 0.0 & 0.0 & ... & 0.0 & 0.0056 & 0.0 \\
\end{bmatrix}
\end{equation*}

\begin{equation*}
c_{i,k} = 
\begin{bmatrix}
1.0 & 0.0 & 1.0 & ... & 4.5826 & 4.899 & 5.3852 \\
2.0 & 1.0 & 0.0 & ... & 4.4721 & 4.5826 & 4.899 \\
3.0 & 2.0 & 1.0 & ... & 4.5826 & 4.4721 & 4.5826 \\
\vdots & \vdots & \vdots & \ddots & \vdots & \vdots & \vdots \\
4.5826 & 4.4721 & 4.5826 & ... & 1.0 & 2.0 & 3.0 \\
4.899 & 4.5826 & 4.4721 & ... & 0.0 & 1.0 & 2.0 \\
5.3852 & 4.899 & 4.5826 & ... & 1.0 & 0.0 & 1.0 \\
\end{bmatrix}
\end{equation*}

\begin{align*}
    EMD_{P_{1},P_{3}} &= \sum_{i,k} \gamma_{i,k}^{*} * c_{i,k} \\
        &= 0.2984
\end{align*}
Finally, 

\[
\Biggl\{
\begin{aligned}
    &EMD_{P_{1},P_{2}} < EMD_{P_{1},P_{3}}, \\
    &SP_{P_{1},P_{2}} > SP_{P_{1},P_{3}}
\end{aligned}
\Biggr.
\]

\clearpage

\subsection*{Euclidean Distance}
The Euclidean Distance between \( P_1 \) and \( P_2 \), here called $ed$ is calculated as:
\begin{align*}
A &= P_1 - P_2 \\
&=
\begin{bmatrix}
\begin{bmatrix}
-1 & 1 & 1 & 1 & -1 \\
1 & 0 & 0 & 1 & 2 \\
1 & -1 & 2 & 0 & -2 \\
0 & 1 & 1 & 1 & 2 \\
0 & 0 & 0 & 0 & -2
\end{bmatrix}, 
\begin{bmatrix}
-1 & 1 & 0 & 0 & 1 \\
-2 & -1 & -1 & -1 & 1 \\
-1 & -1 & -1 & 1 & 0 \\
-2 & -1 & 2 & 2 & 1 \\
1 & -1 & -3 & -1 & 1
\end{bmatrix},
\begin{bmatrix}
-1 & 0 & 1 & 1 & -1 \\
1 & 0 & -2 & -2 & 2 \\
1 & 1 & -2 & 0 & -2 \\
1 & 0 & -2 & -2 & 2 \\
2 & 1 & 2 & 2 & -3
\end{bmatrix}
\end{bmatrix}
\end{align*}

Taking the absolute values:
\begin{align*}
|A| &=
\begin{bmatrix}
\begin{bmatrix}
1 & 1 & 1 & 1 & 1 \\
1 & 0 & 0 & 1 & 2 \\
1 & 1 & 2 & 0 & 2 \\
0 & 1 & 1 & 1 & 2 \\
0 & 0 & 0 & 0 & 2
\end{bmatrix},
\begin{bmatrix}
1 & 1 & 0 & 0 & 1 \\
2 & 1 & 1 & 1 & 1 \\
1 & 1 & 1 & 1 & 0 \\
2 & 1 & 2 & 2 & 1 \\
1 & 1 & 3 & 1 & 1
\end{bmatrix}, 
\begin{bmatrix}
1 & 0 & 1 & 1 & 1 \\
1 & 0 & 2 & 2 & 2 \\
1 & 1 & 2 & 0 & 2 \\
1 & 0 & 2 & 2 & 2 \\
2 & 1 & 2 & 2 & 3
\end{bmatrix}
\end{bmatrix}
\end{align*}

\begin{equation*}
ed_{P1,P2} = \sum |A| = 84
\end{equation*}

Similarly, for \( P_1 \) and \( P_3 \):
\begin{align*}
B &= P_1 - P_3 \\
&=
\begin{bmatrix}
\begin{bmatrix}
-1 & 0 & 1 & 0 & -1 \\
1 & 5 & 1 & 1 & 1 \\
0 & 0 & -3 & -2 & 0 \\
1 & -1 & -1 & -1 & 1 \\
-1 & 0 & 1 & 0 & -1
\end{bmatrix},
\begin{bmatrix}
-1 & 1 & 2 & 1 & 0 \\
0 & 1 & 0 & 0 & 2 \\
1 & 0 & 0 & -2 & 1 \\
0 & 0 & -1 & -1 & 2 \\
0 & 1 & 0 & 1 & 0
\end{bmatrix},
\begin{bmatrix}
-1 & 1 & 3 & 2 & -1 \\
1 & -2 & 1 & 1 & 3 \\
1 & 0 & -2 & 4 & 2 \\
1 & -2 & 1 & 1 & 3 \\
1 & 1 & 3 & 2 & -1
\end{bmatrix}
\end{bmatrix}
\end{align*}

Taking the absolute values:
\begin{align*}
|B| &=
\begin{bmatrix}
\begin{bmatrix}
1 & 0 & 1 & 0 & 1 \\
1 & 5 & 1 & 1 & 1 \\
0 & 0 & 3 & 2 & 0 \\
1 & 1 & 1 & 1 & 1 \\
1 & 0 & 1 & 0 & 1
\end{bmatrix}, 
\begin{bmatrix}
1 & 1 & 2 & 1 & 0 \\
0 & 1 & 0 & 0 & 2 \\
1 & 0 & 0 & 2 & 1 \\
0 & 0 & 1 & 1 & 2 \\
0 & 1 & 0 & 1 & 0
\end{bmatrix},
\begin{bmatrix}
1 & 1 & 3 & 2 & 1 \\
1 & 2 & 1 & 1 & 3 \\
1 & 0 & 2 & 4 & 2 \\
1 & 2 & 1 & 1 & 3 \\
1 & 1 & 3 & 2 & 1
\end{bmatrix}
\end{bmatrix}
\end{align*}

\begin{equation*}
ed_{P1,P3} = \sum |B| = 84
\end{equation*}

So:

\begin{equation*}
ed_{P1,P2} = ed_{P1,P3}
\end{equation*}

\end{document}